# The Role of Weak Interactions in Characterizing Peptide Folding Preferences using a QTAIM Interpretation of the Ramachandran Plot ($\phi$-$\psi$)

**Roya Momen[1], Alireza Azizi[1], Lingling Wang [1], Yang Ping [1], Tianlv Xu[1], Steven R. Kirk[*1] Wenxuan Li[2], Sergei Manzhos[2] and Samantha Jenkins[*1]**

[1]*Key Laboratory of Chemical Biology and Traditional Chinese Medicine Research and Key Laboratory of Resource Fine-Processing and Advanced Materials of Hunan Province of MOE, College of Chemistry and Chemical Engineering, Hunan Normal University, Changsha, Hunan 410081, China*
[2]*Department of Mechanical Engineering, National University of Singapore, Block EA 07-08, 9 Engineering Drive 1, Singapore 117576*

*email: samanthajsuman@gmail.com
*email: steven.kirk@cantab.net

The Ramachandran plot is a potent way to understand structures of biomolecules, however, the original formulation of the Ramachandran plot only considers backbone conformations. We formulate a new interpretation of the original Ramachandran plot ($\phi$-$\psi$) that can include a description of the weaker interactions including both the hydrogen bonds and H---H bonds as a new way to derive insights into the phenomenon of peptide folding. We use QTAIM (quantum theory of atoms in molecules) to interpret the Ramachandran plot. Specifically, we show that QTAIM analysis permits identifying key regions of the Ramachandran plot without the need for massive data sets. A highly non-linear relationship is found between the QTAIM vector-derived interpreted Ramachandran plot and the conventional Ramachandran plot ($\phi$-$\psi$) demonstrating that this new approach is not a trivial coordinate transformation. An investigation of both the backbone and the weaker bonds within the framework of the QTAIM interpreted Ramachandran plot was found to be in line with physical intuition. The least-preferred directions calculated for the hydrogen bonds and H---H bonds were found to coincide with the 'unlikely' regions of the Ramachandran plot.

## 1. Introduction

Recent years have seen advances in the understanding of the chemistry of protein structures and specifically protein folding based on both experimental and theoretical research[1–3]. Ramachandran, Sasisekharan and Ramakrishnan introduced a ground-breaking visualization scheme for protein structures in 1963[4]. The result of which was the creation of the Ramachandran plot ($\phi$-$\psi$) based on the distribution of torsion angles $\varphi$ and $\psi$ of the peptide residues that comprise the backbone[5–7], see **Scheme 1**. The overwhelming success of the Ramachandran plot ($\phi$-$\psi$) is due to the convenience of the universal applicability to peptides due to the huge depth of experience with massive data sets using the Ramachandran plot ($\phi$-$\psi$).

The Ramachandran plot ($\phi$-$\psi$) partitions 'favored', 'allowed' and 'unlikely' regions defined by the $\phi$ and $\psi$ torsion angle values in a protein[8]. Regions where the distribution of the $\varphi$ and $\psi$ angles peaks are defined to be 'favored' or 'allowed'. Conversely, 'unlikely' regions of the $\varphi$ and $\psi$ angles correspond to residues that experience steric clashes and hindrance due to the proximity of the constituent atoms[9,10]. The presence of weaker interactions such as hydrogen bonds[11] and H---H bonds are important for biological[12] and chemical operations[13,14] and are associated with the peptide residues that are indicated by the $\varphi$ and $\psi$ angle values in the 'unlikely' regions due to reduced steric hindrance.

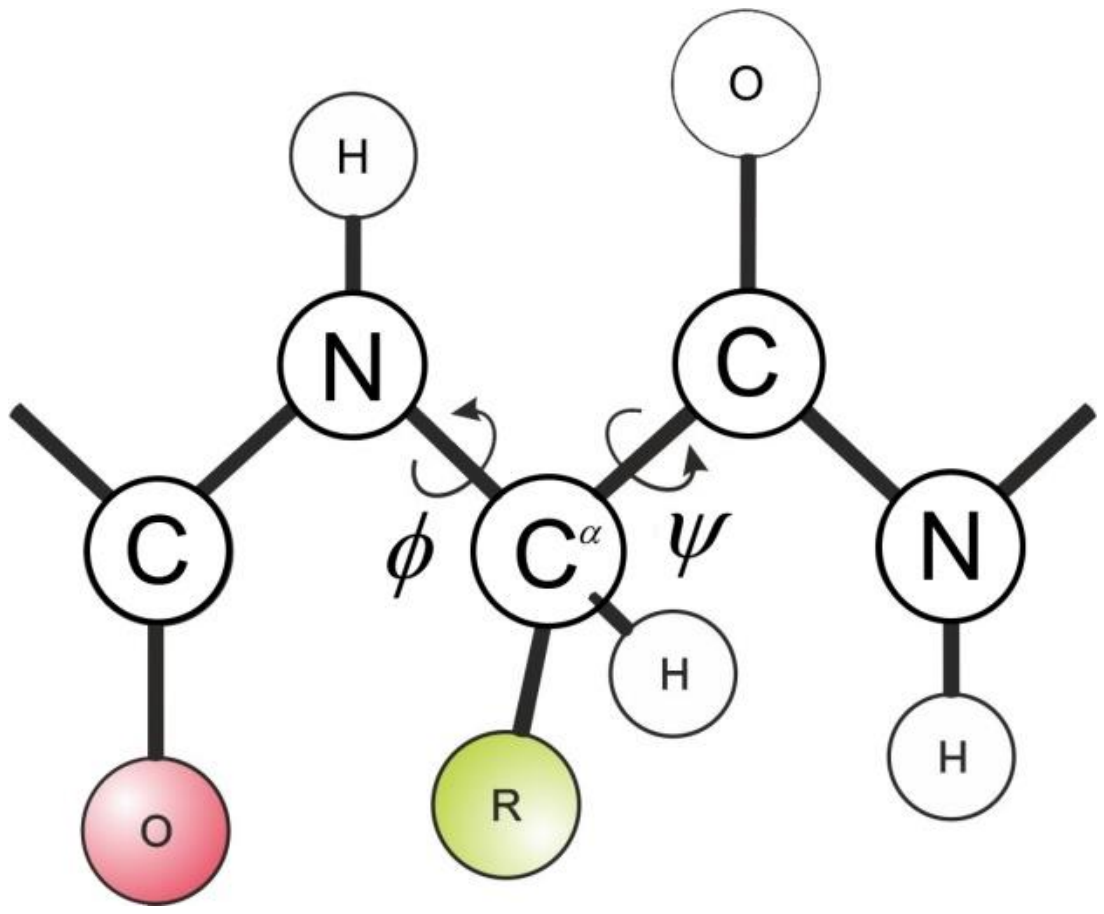

**Scheme 1. Conventional Ramachandran plot ($\phi$-$\psi$) angles $\phi$ and $\psi$ for the peptide backbone.**

As mentioned previously, the Ramachandran plot ($\phi$-$\psi$) is defined only in terms of the evaluation of the main chain conformations i.e. the backbone. The side chain residues however, must also be considered to account for the differences between folding arrangements for the same amino acid sequence within the Ramachandran plot ($\phi$-$\psi$) formalism[15]. In this work we shall only be considering peptides with α-helix structure where the α-helix is defined as a secondary structure of the protein that is stabilized by hydrogen bonds between main chain amide and carbonyl group[16,17]. Therefore, hydrogen bonding[18–21] and H---H bonds[22] clearly play an important role in the stabilization of the peptide and typically occur between carbonyl and amide groups separated by four amino acids within the α-helix. Additionally, the important role of hydrogen bonding cooperativity[23–25] has been discussed recently to understand the secondary structures of peptides [26,27].

The main goal of this work is to formulate a new interpretation of the original Ramachandran plot ($\phi$-$\psi$) that can include a description of the weaker interactions, including both the hydrogen bonds and H---H bonds as a new way to derive insights into the phenomenon of peptide folding[28–31].

In this exploratory work we investigate the extent to which analogous 'favored', 'allowed' and 'unlikely' regions can be created from the quantum theory of atoms in molecules (QTAIM) [32] without the need for massive data sets of structures. Instead, we will use indications from the vector character of QTAIM determined at the bond critical points. We undertake this investigation with a series of five lycosin conformers. Lycosin has in fact recently attracted increasing attention as it has been shown to be a promising candidate for the inhibition of tumor cell growth in vitro and in vivo[33 34]. The mechanism of this action is still under investigation. Because peptide conformations critically affect the cell penetrating and cytotoxic functions, theoretical insights gained based on the variation of their topology are likely to be useful.

## 2. Theory and Methods

### *2.1 The QTAIM BCP scalar descriptors; ellipticity ε and stiffness 𝕊*

We will use QTAIM[32] to identify critical points in the total electronic charge density distribution $\rho(\mathbf{r})$ by analyzing $\nabla\rho(\mathbf{r})$. These critical points can further be divided into four types of topologically stable critical points according to the set of ordered eigenvalues $\lambda_1 < \lambda_2 < \lambda_3$, with corresponding eigenvectors $\underline{\mathbf{e}}_1$, $\underline{\mathbf{e}}_2$, $\underline{\mathbf{e}}_3$ of the Hessian matrix. The Hessian of the total electronic charge density $\rho(\mathbf{r})$ is defined as the matrix of partial second derivatives with respect to the spatial coordinates. These critical points are labeled using the notation ($R$, ω) where $R$ is the rank of the Hessian matrix, the number of distinct non-zero eigenvalues and ω is the signature (the algebraic sum of the signs of the eigenvalues); the (3, -3) [nuclear critical point (*NCP*), a local maximum generally corresponding to a nuclear location], (3, -1) and (3, 1) [saddle points, called bond critical points (*BCP*) and ring critical points (*RCP*), respectively] and (3, 3) [the cage critical points (*CCP*)]. The presence of *CCP*s is associated with a resistance of a structure to being crushed [35]. In the limit that the forces on the nuclei become vanishingly small, an atomic interaction line (AIL)[36] becomes a bond-path, although not necessarily a chemical bond[37]. The complete set of critical points together with the bond-paths of a molecule or cluster is referred to as the molecular graph. For molecules and clusters the Poincaré-Hopf relation is expressed as:

$$n - b + r - c = 1\,, \qquad \textbf{(1)}$$

where the four types of critical points; *n*, *b*, *r*, and *c* are given by the numbers of *NCPs*, *BCPs*, *RCPs*, and *CCPs*, respectively. The sum of the number of these critical points for a given molecular graph is referred to as the topological complexity $\sum_{brc}$. In this investigation we have closed-shell *BCP*s, H--O *BCP*s, H---H *BCP*s and backbone shared-shell *BCP*s that comprise the peptide backbone where the Laplacian $\nabla^2\rho(\mathbf{r}) > 0$ for the closed-shell *BCP*s and the Laplacian $\nabla^2\rho(\mathbf{r}) < 0$ for backbone shared-shell *BCP*s.

The ellipticity, ε provides the relative accumulation of $\rho(\mathbf{r_b})$ in the two directions perpendicular to the bond-path at a *BCP*, defined as $\varepsilon = |\lambda_1|/|\lambda_2| - 1$ where $\lambda_1$ and $\lambda_2$ are negative eigenvalues of the corresponding Hessian matrix at the *BCP*[32].

A degree of covalent character is indicated for $H(\mathbf{r}_b) < 0$ [38], where the total local energy density $H(\mathbf{r}_b) = G(\mathbf{r}_b) + V(\mathbf{r}_b)$ where $G(\mathbf{r}_b)$ and $V(\mathbf{r}_b)$ are the local kinetic and virial energy densities [32], respectively. Previously, one of the current authors used $H(\mathbf{r}_b) < 0$ to quantify the degree of covalent character present in the hydrogen-bonding in ice Ih to provide a quantitative explanation of the unusually high strength of ice Ih hydrogen-bonds[25]. This result was explained in terms of coupling i.e., cooperativity between the hydrogen-bonds and adjoining (σ) O-H bond paths that share an H *NCP* causing covalent bond character to 'leak' from the O-H *BCP*s to the hydrogen-bond *BCP*s[25]. Throughout this work, we use the notation H--O *BCP* and H---O *BCP* for hydrogen bond *BCP*s with $H(\mathbf{r}_b) < 0$ and $H(\mathbf{r}_b) > 0$ respectively, all the H---H *BCP*s possess $H(\mathbf{r}_b) > 0$.

We define the stiffness, $\mathbb{S} = |\lambda_2|/\lambda_3$, where the larger the value of $\lambda_3$ the 'stiffer' the bond-path is, explained in terms of the contraction of $\rho(\mathbf{r_b})$ away from the *BCP* toward the nuclei. The presence of the eigenvalue, $\lambda_2$ accounts for the direction of the preferred accumulation of charge density: $\underline{\mathbf{e}}_2$ and therefore the direction of the π-bond[38].

### *2.2 The QTAIM interpreted Ramachandran plot ($\beta_\phi$,$\beta_\psi$)*

In recent investigations, the directional properties derived at from the total electronic charge density $\rho(\mathbf{r_b})$ a *BCP* do not always move in accordance with the nuclei[38]. Instead, the resistance of the total electronic charge density $\rho(\mathbf{r_b})$ to the deformation of the nuclear skeleton may depend on the electronic state or nature of the nuclei involved. To describe the deformation of a bond-path using the QTAIM we recently defined the {$\underline{\mathbf{e}}_1$, $\underline{\mathbf{e}}_2$, $\underline{\mathbf{e}}_3$} bond-path framework as the set of orthogonal $\underline{\mathbf{e}}_1$, $\underline{\mathbf{e}}_2$ and $\underline{\mathbf{e}}_3$ eigenvectors of the Hessian matrix for the two bonded nuclei that comprised the torsional bond-path[39], see **Scheme 2** which explains the definition of the response *β*.

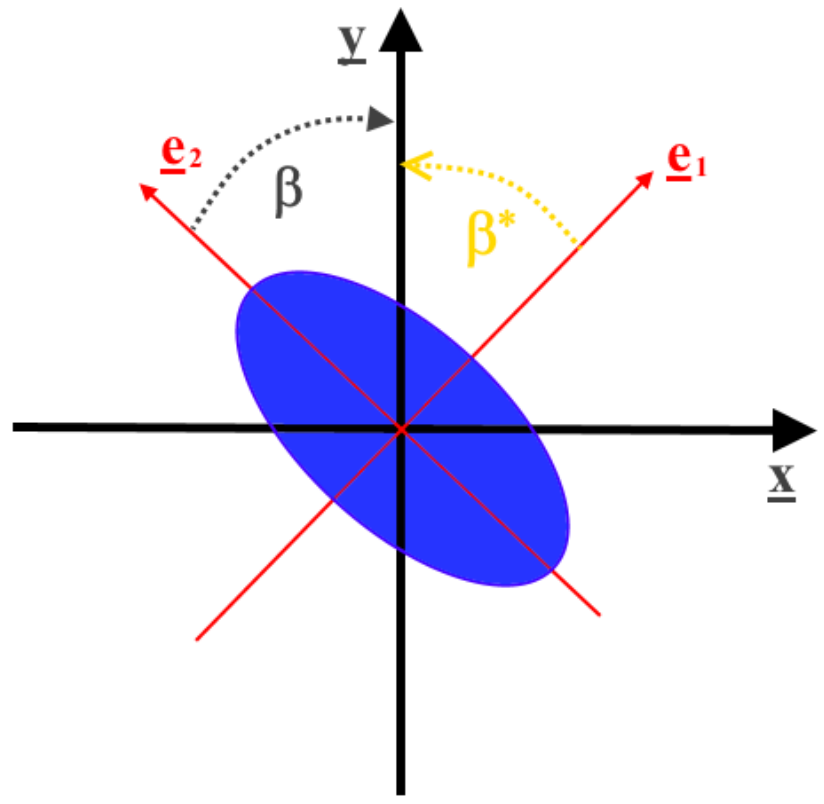

**Scheme 2**. The {$\underline{\mathbf{e}}_1$, $\underline{\mathbf{e}}_2$, $\underline{\mathbf{e}}_3$} bond-path framework is shown perpendicular to the $\underline{\mathbf{e}}_3$ eigenvector. The solid blue ellipse represents the cross-section of a bond-path with the most and least preferred directions of the total charge density distribution $\rho(\mathbf{r})$ indicated by the $\underline{\mathbf{e}}_2$ and $\underline{\mathbf{e}}_1$ eigenvectors respectively. The most ($\beta$) and least ($\beta^*$) preferred eigenvector responses are given by $\beta$, $\beta^*$ = arccos($\underline{\boldsymbol{v}}\cdot\underline{\boldsymbol{n}}$), where $\underline{\boldsymbol{v}} = \underline{\mathbf{e}}_2$, $\underline{\boldsymbol{v}} = \underline{\mathbf{e}}_1$ for $\beta$ and $\beta^*$ indicated with black and orange fonts respectively.

The response $\beta$ is defined as:

$$\beta = arc\cos(\underline{\mathbf{e}}_2 \cdot \underline{\mathbf{y}}) \qquad \textbf{(2)}$$

Where $\underline{\mathbf{y}}$ is a reference vector of unit length along the **Y** axis. Recently, we demonstrated that the eigenvectors $\underline{\mathbf{e}}_2$ and $\underline{\mathbf{e}}_1$ were the most and least preferred directions of torsion respectively using the photo-isomerization of the retinal chromophore subject to a torsion -$\phi$ or +$\phi$[40]. In this recent work the response $\beta$ of the total electronic charge density $\rho(\mathbf{r}_b)$ to torsion about a pivot-*BCP* was introduced for different electronic states of the fulvene molecular graph[41]. Therefore, the response $\beta$ relates to the degree of detachment of the {$\underline{\mathbf{e}}_1$, $\underline{\mathbf{e}}_2$, $\underline{\mathbf{e}}_3$} framework from the containing nuclear skeleton. Smaller values of $\beta$ indicate a greater degree of *alignment* of the {$\underline{\mathbf{e}}_1$, $\underline{\mathbf{e}}_2$, $\underline{\mathbf{e}}_3$} framework with the containing nuclear skeleton or peptide backbone. Conversely, larger values of $\beta$ indicate a greater degree of *detachment* of the {$\underline{\mathbf{e}}_1$, $\underline{\mathbf{e}}_2$, $\underline{\mathbf{e}}_3$} from the containing nuclear skeleton. For this work, unlike the earlier work on fulvene, we are not concerned with applied torsions α to the nuclear skeleton and so cannot directly associate the response $\beta$ with the presence of double bond or single bond character. Instead, we can associate low values of the ellipticity ε with a greater sensitivity of $\beta$ to changes in a molecular graph and conversely, higher values of the ellipticity ε with a reduced sensitivity of $\beta$. In this work we will refer to the response $\beta$ as the most preferred (eigenvector) to distinguish it from the least preferred response $\beta^*$:

$$\beta^* = arc\cos(\underline{\mathbf{e}}_1 \cdot \underline{\mathbf{y}}) \qquad \textbf{(3)}$$

In earlier work we showed for the axel bond that linked the two rings in biphenyl that there was a symmetrical dependence on the variation of the $\underline{\mathbf{e}}_1$ defined basin path set area with an applied torsion α[42]. No such symmetrical dependency was found for the $\underline{\mathbf{e}}_2$ defined basin path set area, i.e. the dependency was asymmetric due to the existence of a preferred sense of torsion for the $\underline{\mathbf{e}}_2$ defined basin path set area.

For the closed-shell *BCP*s we chose the stiffness $\mathbb{S}$ in favor of the ellipticity ε because, unlike for the backbone shared-shell *BCP*s, the ellipticity ε does not serve to act as a classifier on the basis of high, i.e. > 0.25 (double bond character) or low values (single bond character). The stiffness $\mathbb{S}$ for the closed-shell *BCP*s however, does indicate a resistance to undergo torsion, where high values of the stiffness $\mathbb{S}$ will not readily be distorted and should correspond to values of $\beta$ close to 0º. The converse being true for low values of the stiffness $\mathbb{S}$, where we expect values of $\beta$ to be close to the maximum value of $\beta$ = 90º.

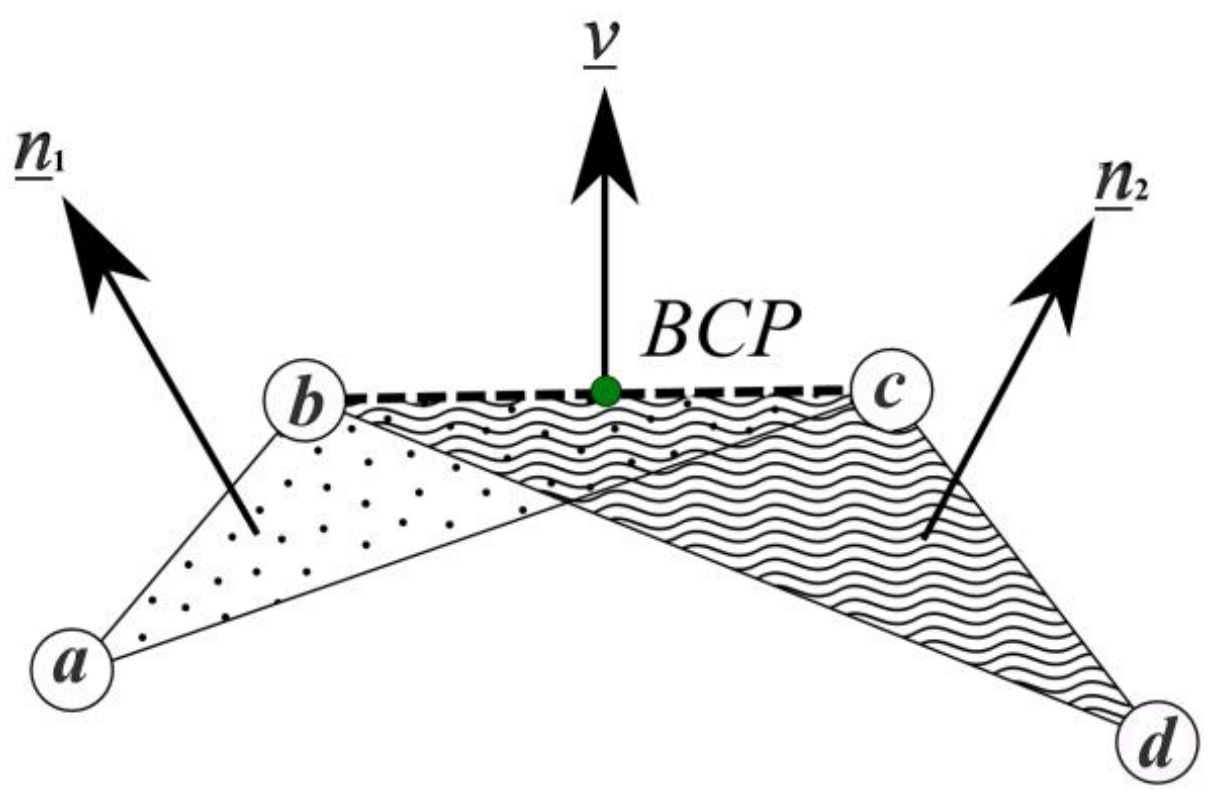

**Scheme 3**. The schematic for defining the QTAIM analogous Ramachandran angles ($\phi$-$\psi$), i.e. the responses ($\beta_{\phi}$,$\beta_{\psi}$) and ($\beta^{*}_{\phi}$,$\beta^{*}_{\psi}$), for the closed-shell H--O/H---O and H---H *BCP*s. The QTAIM most preferred response ($\beta_{\phi}$,$\beta_{\psi}$) and the least preferred response ($\beta^{*}_{\phi}$,$\beta^{*}_{\psi}$) are defined using a vector $\underline{\boldsymbol{v}}$ and unit-length plane normal vectors $\underline{\boldsymbol{n}}_{\mathbf{1}}$ and $\underline{\boldsymbol{n}}_{\mathbf{2}}$ in planes defined by atoms *a*, *b*, *c* and *b*, *c*, *d* respectively. The plane normal vectors $\underline{\boldsymbol{n}}_{\mathbf{1}}$ and $\underline{\boldsymbol{n}}_{\mathbf{2}}$ are used as reference directions analogously to the unit vector $\underline{\mathbf{y}}$ defined in **Scheme 2**, see equation **(4)** and equation **(5)**. We select atoms *a* and *d*, the closest backbone shared-shell bonded atoms to atoms *b* and *c* respectively that comprise the *b*-*c* bond-path of the *b*-*c BCP*. If the *a*-*b* bond-path length is greater than *c*-*d* bond-path length, the order of the atom labels *a*-*b*-*c*-*d* is reversed.

The creation of the conventional Ramachandran plot ($\phi$-$\psi$) uses *torsion* angles ($\phi$-$\psi$) of the backbone nuclear skeleton, therefore it seems to be worth pursuing the idea of the response $\beta$ that also measures the similar *angular* detachment of the {$\underline{\mathbf{e}}_{\mathbf{1}}$, $\underline{\mathbf{e}}_{\mathbf{2}}$, $\underline{\mathbf{e}}_{\mathbf{3}}$} framework to create a QTAIM interpretation of the Ramachandran plot ($\phi$-$\psi$). For closed-shell *BCP*s we use the same ordering conventions used to construct Ramachandran angles $\phi$ and $\psi$, replacing the atomic geometry reference vector $\underline{\mathbf{y}}$, see **Scheme 2**, **equation (2)** and **equation (3)**, with a unit reference vector $\underline{\boldsymbol{n}}$. This vector $\underline{\boldsymbol{n}}$ is derived from the local atomic geometry, specifically the normal vector to a plane through three atomic positions, see **Scheme 3**. For the closed-shell *BCP*s the most preferred response ($\beta_{\phi}$,$\beta_{\psi}$) is defined as:

$$\beta_{\phi},\beta_{\psi} = \arccos(\underline{\boldsymbol{v}}\cdot\underline{\boldsymbol{n}}), \qquad \text{where } \underline{\boldsymbol{v}} = \underline{\mathbf{e}}_{\mathbf{2}} \qquad \textbf{(4)}$$

the least preferred response ($\beta^{*}_{\phi}$,$\beta^{*}_{\psi}$) is defined as:

$$\beta^{*}_{\phi},\beta^{*}_{\psi} = \arccos(\underline{\boldsymbol{v}}\cdot\underline{\boldsymbol{n}}), \qquad \text{where } \underline{\boldsymbol{v}} = \underline{\mathbf{e}}_{\mathbf{1}} \qquad \textbf{(5)}$$

Where $\underline{\boldsymbol{n}} = \underline{\boldsymbol{n}}_1$ and $\underline{\boldsymbol{n}} = \underline{\boldsymbol{n}}_2$ for $\phi$ and $\psi$ respectively. For the backbone shared-shell-*BCP*s will also use equation **(4)** and equation **(5)** but with $\underline{\boldsymbol{n}} = \underline{\boldsymbol{n}}_1$, $\underline{\boldsymbol{n}}_2$ directly corresponding to the plane normal vectors for groups of three successive atoms for the Ramachandran angles $\phi$ and $\psi$ respectively, see **Scheme 4**.

Although the responses ($\beta_\phi$,$\beta_\psi$) and ($\beta^*_\phi$,$\beta^*_\psi$) are derived from the total charge density distribution $\rho(\mathbf{r_b})$ at the *BCP* we can directly compare the responses ($\beta_\phi$,$\beta_\psi$) and ($\beta^*_\phi$,$\beta^*_\psi$) from the closed-shell H--O/H---O and H---H *BCP*s with that of the backbone shared-shell *BCP*s. This is because the definition of the responses ($\beta_\phi$,$\beta_\psi$) and ($\beta^*_\phi$,$\beta^*_\psi$) uses the local atomic geometry which is independent of QTAIM, see the vector $\underline{\mathbf{v}}$ in equation **(2)**, equation **(3)** and **Scheme 3** and **Scheme 4**.

In this work, we have a limited data set; only five lycosin peptides and consequently the resulting QTAIM interpretation of the Ramachandran plot ($\phi$-$\psi$) will not contain sufficient data points to map, with confidence, the *boundaries* of 'favored', 'allowed' and 'unlikely' regions. From QTAIM it should be possible to explicitly create the 'favored', 'allowed' and 'unlikely' regions to form an analogous Ramachandran plot.

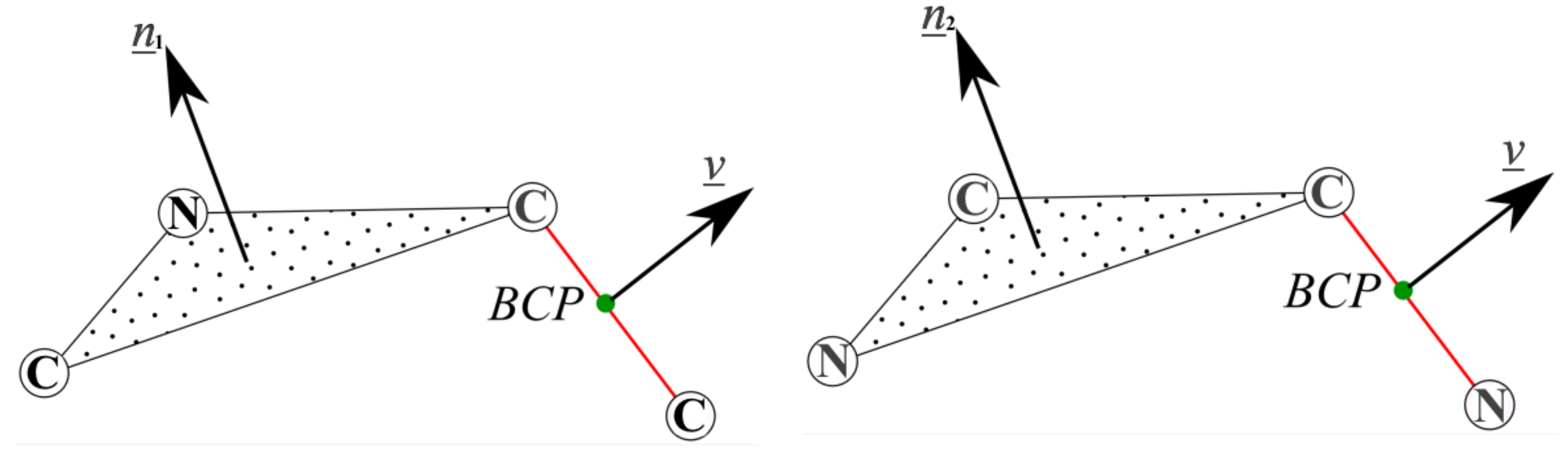

**(a)** **(b)**

**Scheme 4.** The definition of the QTAIM-interpreted Ramachandran angles $\beta_\phi$ and $\beta_\psi$ for the backbone shared-shell *BCPs* are provided in sub-figures **(a)** and **(b)** respectively. In sub-figure **(a)** the repeating sequence of these angles $\beta_\phi$, $\beta_\psi$ is generated as we move along the backbone, analogous to the Ramachandran $\phi$-angle (C:carbon, N:nitrogen, C:**carbon**, C:**carbon**) and in sub-figure **(b)** the Ramachandran $\psi$-angle (N:nitrogen, C:carbon, C:**carbon**, N:**nitrogen**), using the associated $\underline{\boldsymbol{n}}_1$ and $\underline{\boldsymbol{n}}_2$ plane normal vectors, see equation **(4)** and equation **(5)**. The atom type sequences used correspond to those used in the definition of the conventional Ramachandran plot ($\phi$-$\psi$), see **Scheme 1 and Scheme 3**.

## 3. Computational Details

The initial structures of the five lycosin peptide conformers were generated by PEP-FOLD[43–45]. The lowest-energy structures from PEP-FOLD were used as the initial structures for MD simulations; we confirmed that the lowest-energy PEP-FOLD structures were also lowest-energy in DFTB simulations. The MD simulations were performed at 300K with the CHARMM force field[46–48] in the NAMD code[49] for 10ns starting from the lowest-energy PEP-FOLD generated structure. Visually different conformers were selected along the trajectory for optimization with the self-consistent charge density functional tight binding (SCC-DFTB) method[50]. The SCC-DFTB calculations were done with the DFTB+ code[51]. The 3ob-3-1 parameter set was used using the DFTB3 capability[52,53]. The dispersion interactions were modeled with the UFF scheme[54]. Single-point DFT calculations were then performed on DFTB-optimized structures using the Gaussian 09 software[55]. The ωB97XD exchange correlation functional[56] was used to account for any long-range interactions, with the 6-31+g[57] basis set. For the purpose of this work, the density calculations were performed in vacuum, therefore MD (whose sole purpose is to help identify candidate local minima structure) calculations were also performed in vacuum and DFTB and DFT calculations were done on neutral molecules without protonation/deprotonation of amino and carboxylic groups (i.e. –COOH and $-NH_2$ were used rather than $-COO^-$ and $-NH_3^+$). The QTAIM analysis was performed with the AIMAll[58] suite. AIMAll constructs total charge density distributions $\rho(\mathbf{r})$ based on orbital coefficients imported from Gaussian 09: in principle, any method providing total charge density distributions $\rho(\mathbf{r})$ would be suitable for QTAIM analysis. DFT or its approximations such as DFTB place a wide range of sizes of biologically relevant peptide structures within the limits of current or near-future computational resources. In our case, the system size was amenable to DFT calculations in Gaussian and DFT total charge density distributions $\rho(\mathbf{r})$ were therefore used.

## 4. Results and discussions

In this section, we provide a listing of all the figures, tables and Supplementary Materials before the discussion in section 4.1 and section 4.2. The molecular graphs corresponding to the five lycosin peptide structures, lycosin-01, lycosin-02, lycosin-03, lycosin-04 and lycosin-05 are provided in **Figure S1** of the **Supplementary Materials S1**. The summary of the QTAIM topologies of the lycosin-01, lycosin-02, lycosin-03, lycosin-04 and lycosin-05 molecular graphs along with the corresponding relative energies ΔE are given in **Table 1**. Plots of the combined response $\beta_\phi$ versus $\beta_\psi$ and response $\beta^*_\phi$ versus $\beta^*_\psi$, with the distribution of stiffness $\mathbb{S}$ of the H--O *BCP*s, H---O *BCP*s and H---H *BCPs*, for the lycosin-01, lycosin-02, lycosin-03, lycosin-04 and lycosin-05 molecular graphs are shown in **Figure 2**. The plots of the response $\beta_\phi$ versus $\beta_\psi$ with the distribution of ellipticity ε of the backbone shared-shell *BCP*s for the lycosin-01, lycosin-02, lycosin-03, lycosin-04 and lycosin-05 molecular graphs are presented in **Figure 3**. The definition of the response $\beta$ and $\beta^*$ is provided in **Scheme 2**. The QTAIM-interpreted Ramachandran angles $\phi$ and $\psi$ that is; $\beta_\phi$, $\beta_\psi$, for both the closed-shell *BCP*s and backbone shared-shell *BCP*s are provided in **Scheme 3** and **Scheme 4 respectively** and in **Figure 4**. The Ramachandran plots ($\phi$-$\psi$) for the five lycosin peptide conformers are provided in **Supplementary Materials S2 and Table S2**. Plots of the responses ($\beta_\phi$.,$\beta_\psi$) and ($\beta^*_\phi$,$\beta^*_\psi$) versus the distribution of stiffness $\mathbb{S}$ of the closed-shell *BCP*s for each of the five lycosin molecular graphs are provided as separate plots in **Supplementary**

**Materials S3**. Plots of the responses $(\beta_{\phi.},\beta_{\psi})$ and $(\beta^{*}_{\phi},\beta^{*}_{\psi})$ versus the distribution of stiffness $\mathbb{S}$ of the backbone shared-shell *BCP*s for each of the five lycosin molecular graphs are provided as separate plots in **Supplementary Materials S4**.

### *4.1. The QTAIM topology and the closed-shell BCPs*

The conventional Ramachandran plot ($\phi$-$\psi$) does not include a consideration of the closed-shell *BCP*s, therefore in this section we consider the closed-shell *BCP*s, see **Supplementary Materials S2**.

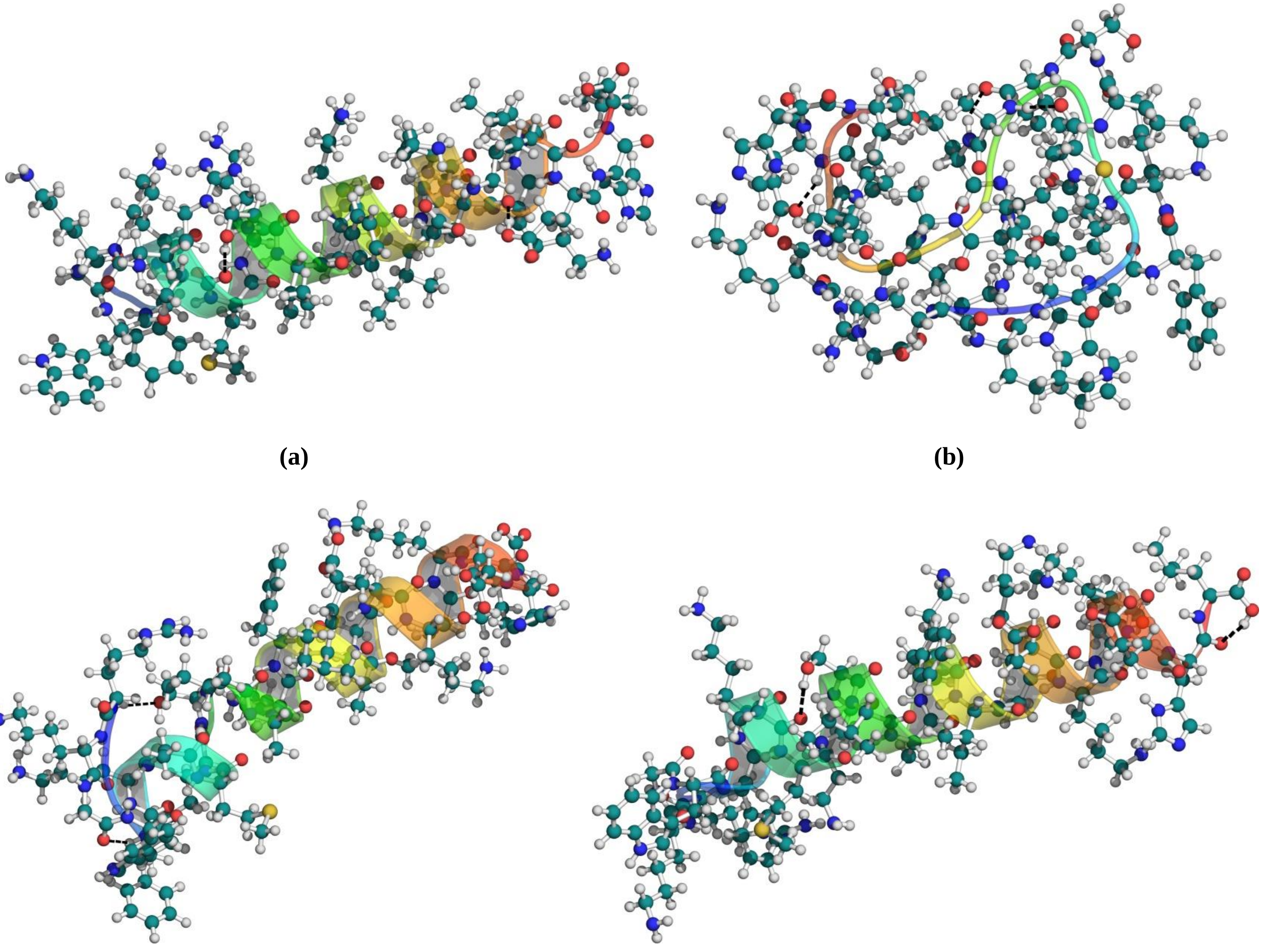

**(a)** **(b)**

**(c)** **(d)**

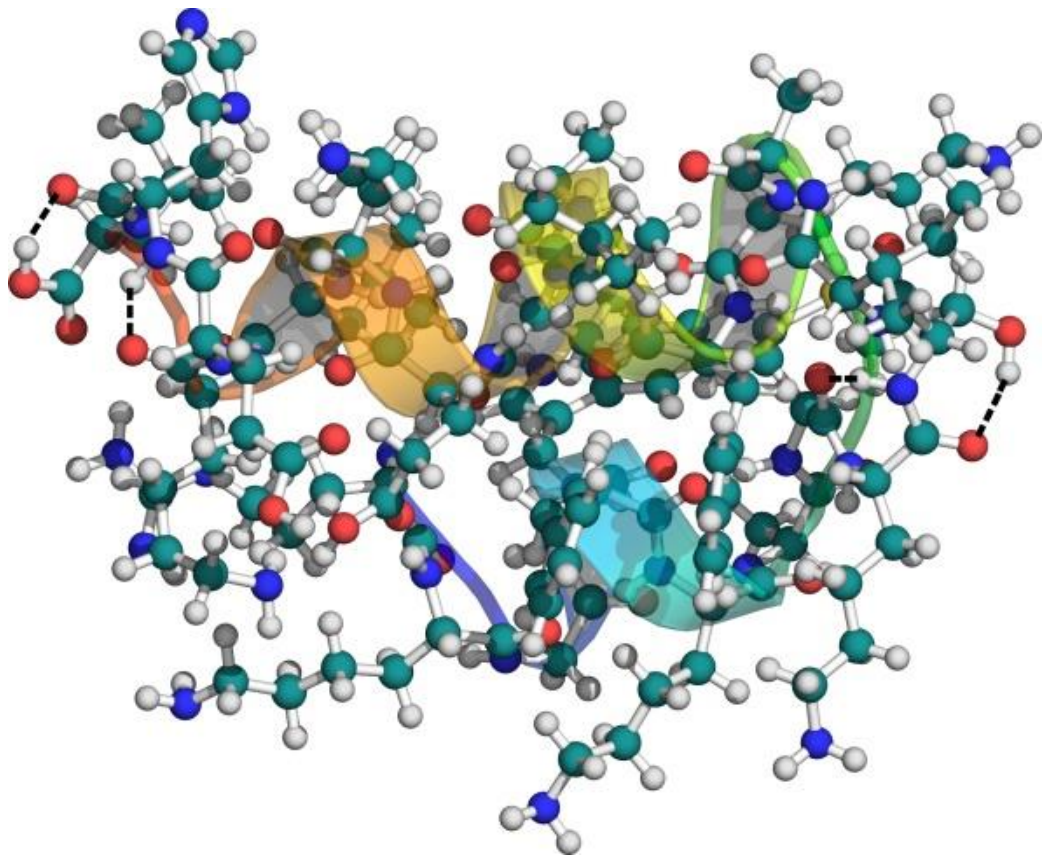

**(e)**

**Figure 1**. The ball and stick renderings of the lycosin-01, lycosin-02, lycosin-03, lycosin-04 and lycosin-05 conformers are presented in sub-figures **(a-e)** respectively. The carbon, hydrogen, oxygen, nitrogen and sulphur atoms are indicated by the green, white, red, blue and yellow spheres respectively. The colored ribbon is superimposed on the structures to represent the α-helix where blue and red represent the start and termination of the RKGWFKAMKSIAKFIAKEKLKEHL amino acid sequence. Hydrogen bond *BCP* bond-paths possessing a degree of covalent character, that is total local energy density $H(\mathbf{r_b}) < 0$ are indicated by the thick black dashed lines.

For the five lycosin conformers there are differences in the QTAIM topologies, i.e. the numbers of *BCP*s (*b*), *RCP*s (*r*) and *CCP*s (*c*) as well as the sum of these critical points for each conformer, referred to as the topological complexity $\sum_{brc}$, see **Table 1**. The number of *CCP*s comprising each molecular graph is the reason that the tube-like appearance corresponding to the α-helix is visible for lycosin-01, lycosin-03, lycosin-04 and lycosin-05 but not for lycosin-02, the most energetically unstable structure, see **Table 1**. This is because the lowest numbers of *CCP*s comprising the lycosin-02 molecular graph indicate that the structure is the most crumpled. For lycosin-05, the most energetically stable structure, the linear helix structure is folded by the particular positioning of 'sticky' hydrogen bond H--O *BCP*s i.e. possessing a degree of covalent character $H(\mathbf{r_b}) < 0$, see **Figure 1(e)**.

**Table 1**. The solution sets of the Poincaré-Hopf relation ($n$ = 427), where *b*, *r* and *c* represent the numbers of *BCP*s, *RCP*s and *CCP*s, see equation **(1)**. The topological complexity $\Sigma_{brc}$ and relative energy ΔE (computed with DFT/ωB97XD/6-31+g) for the molecular graphs of the five conformers of lycosin, see **Figure S1**. The α-helix structure lycosin peptide with the twenty four amino acid sequence RKGWFKAMKSIAKFIAKEKLKEHL[34,59].

| **Molecular graph** | *b* | *r* | *c* | $\Sigma_{brc}$ | ΔE(au) |
|---|---|---|---|---|---|
| Lycosin-05 | 593 | 219 | 52 | 864 | 0.0000 |
| Lycosin-03 | 574 | 194 | 46 | 814 | 0.0091 |
| Lycosin-04 | 564 | 185 | 47 | 796 | 0.0297 |
| Lycosin-01 | 550 | 170 | 46 | 766 | 0.0399 |
| Lycosin-02 | 585 | 202 | 43 | 830 | 0.0433 |

In this section we focus on the preferred response angles ($\beta_\phi$, $\beta_\psi$), for the H--O *BCP*, H---O *BCP* and H---H *BCP*s closed-shell *BCP*s, see **Figure 2** colored mapped to the stiffness $\mathbb{S}$, respectively. It can be seen that in each case that the form of the plots for the most preferred response ($\beta_\phi$,$\beta_\psi$) and the least preferred response ($\beta^*_\phi$,$\beta^*_\psi$) are characteristically different, see equation (**4**) and equation (**5**) respectively, see **Figure 2**. The distribution of data points for most preferred response ($\beta_\phi$,$\beta_\psi$) and least preferred response ($\beta^*_\phi$,$\beta^*_\psi$) is complementary for the H--O/H---O *BCP*s and H---H *BCP*s, i.e. there are characteristic regions for ($\beta_\phi$,$\beta_\psi$) and ($\beta^*_\phi$,$\beta^*_\psi$), that can be seen by comparing sub-figure (**a**) and (**b**) for each of **Figure 2** and **Supplementary Materials S3**. For the H--O *BCP*s and H---O *BCP*s there is a clustering of the least preferred response ($\beta^*_\phi$,$\beta^*_\psi$) in the vicinity of $\beta^*_\phi$ = 0.0º. This can be explained by the inherent asymmetry in the construction of the ($\beta_\phi$,$\beta_\psi$) and ($\beta^*_\phi$,$\beta^*_\psi$); the bond-path length of the first nearest neighbor backbone shared-shell *BCP* used in the construction of $\beta_\phi$ is shorter than that for $\beta_\psi$, see **Scheme 3**. All of the H--O *BCP*s, i.e. the *BCP*s that possess a degree of covalent character on the basis of $H(\mathbf{r_b}) < 0$ possess greater values of the stiffness $\mathbb{S}$ than do the H---O *BCP*s or H---H *BCP*s that possess $H(\mathbf{r_b}) > 0$, in line with expectations from physical intuition, see the square symbols in sub-figure (**a**) and (**b**) for each of **Figure 2** and **Supplementary Materials S3**. Similarly, the stiffness $\mathbb{S}$ values of the purely electrostatic H---O *BCP*s are generally greater than those of the H---H *BCP*s. The ($\beta_\phi$,$\beta_\psi$) values for the H--O *BCP*s are generally positioned closer to the $\beta_\phi$ = 0.0º and $\beta_\psi$ = 0.0º than are the corresponding ($\beta^*_\phi$,$\beta^*_\psi$) values for the H--O *BCP*s. Values of $\beta$ closer to 0.0º indicate a greater degree of *alignment* of the {$\underline{\mathbf{e}}_1$, $\underline{\mathbf{e}}_2$, $\underline{\mathbf{e}}_3$} framework with the containing nuclear skeleton or backbone, see the theory section 2.3 for further explanation. For the purely electrostatic H---O *BCP*s the ($\beta_\phi$,$\beta_\psi$) plots there is a tendency for the H---O *BCP*s with higher stiffness $\mathbb{S}$ values, indicated in red, to be located closer to $\beta_\psi$ = 0.0º than the H---O *BCP*s with lower stiffness $\mathbb{S}$ values, shown in blue in the electronic version. The converse is true for the corresponding ($\beta^*_\phi$,$\beta^*_\psi$) values for the H---O *BCP*s.

The ($\beta_\phi$,$\beta_\psi$) and ($\beta^*_\phi$,$\beta^*_\psi$) values for the H---H *BCP*s are distributed rather more symmetrically than is the case for the H--O *BCP*s and H---O *BCP*s, compare (**c**) and (**d**) for each of **Figure 2** This is because the nearest neighbors are generally backbone shared-shell C-H *BCP*s at both ends of the H---H *BCP*. The preference for ($\beta_\phi$,$\beta_\psi$) = ±90.0º values can be explained by the close proximity of an *RCP* to each H---H *BCP* where the $\underline{\mathbf{e}}_2$ eigenvector and therefore also $\beta_\phi$ or $\beta_\psi$ that indicates the preferred direction of electronic charge density accumulation $\rho(\mathbf{r_b})$ connects a *BCP* and nearest *RCP*. The discussion of the backbone shared-shell *BCP*s in the next section indicates that the conventional Ramachandran plot ($\phi$-$\psi$) suggests a degree of beta-sheet character for the lycosin-02 peptide, where hydrogen bonds of beta-sheets are considered to be slightly stronger than those found in α-helices[60–62], see also **Supplementary Materials S1**. Note, we should not confuse the term 'beta-sheet' with the response $\beta$. We suggest that the form of distribution of the H--O/H---O *BCP* data of the preferred response ($\beta_\phi$,$\beta_\psi$) is characteristic of structures containing α-helix structures.

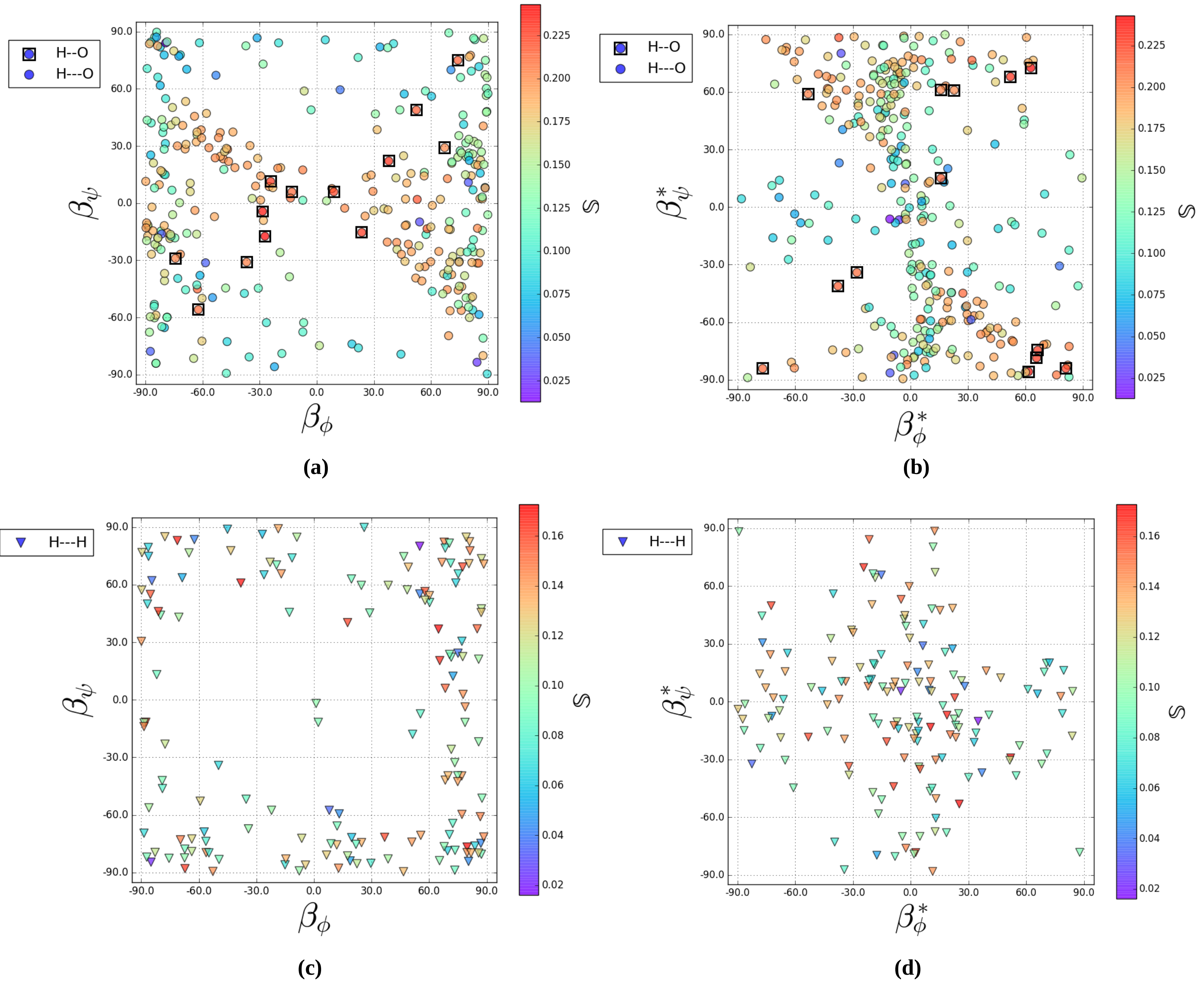

**Figure 2.** Plots of the (preferred) response $\beta_\phi$ versus $\beta_\psi$ with the distribution of stiffness $\mathbb{S}$ of the H--O *BCP*s, H---O *BCP*s for the molecular graph of lycosin-01, lycosin-02, lycosin-03, lycosin-04 and lycosin-05 are shown in sub-figure **(a)**. The corresponding quantity for the least preferred responses ($\beta^*_\phi$,$\beta^*_\psi$) of the H--O *BCP*s, H---O *BCP*s are given in sub-figure **(b)**. The data points surrounded by squares correspond to values of the total local energy density $H(\mathbf{r}_\mathbf{b}) < 0$. Plots of the responses $\beta_\phi$ versus $\beta_\psi$ with the distribution of stiffness $\mathbb{S}$ of the H---H *BCPs* are shown in sub-figures **(c)**. The corresponding quantity for the least preferred responses ($\beta^*_\phi$,$\beta^*_\psi$) of the H---H *BCP*s are given in sub-figure **(d)**. The responses represent the QTAIM interpretation of the Ramachandran plot ($\phi$-$\psi$) and indicate the most and least preferred eigenvector responses to structural distortions, see **Scheme 3** for the explanation of the construction of the ($\beta_\phi$,$\beta_\psi$) and ($\beta^*_\phi$,$\beta^*_\psi$) for closed-shell *BCP*s.

*4.2. Towards a QTAIM interpretation of the Ramachandran plot ($\phi$-$\psi$) for peptide folding*

Another weakness of the Ramachandran plot ($\phi$-$\psi$), apart from not considering weak interactions, see section 4.1, is the inability to correctly characterize the secondary structure. This can be seen by comparing the sense of rotation of the helix in each structure and the existence of data points in the left-hand α-helix region of the Ramachandran plot ($\phi$-$\psi$), see **Supplementary Materials S2**. According to the Ramachandran plot ($\phi$-$\psi$) lycosin-01 and lycosin-02 are the only peptide structures that comprise left handed α-helices, see the orange squares located in the left-hand α-helix region in **Figure S1(a)** and **Figure S1(b)** of **Supplementary Materials S2**. Inspection of **Figure 1** however, shows that the lycosin-01 structure comprises only a right handed α-helix and that there is no α-helix at all present for lycosin-02, showing the inconsistency of the Ramachandran plot ($\phi$-$\psi$). The lycosin-05 conformer contains a portion AKFIAKEKLKEHL of the backbone that has left-handed turns, see **Figure 1(e)** but there are no data points in the left hand α-helix region of the Ramachandran plot ($\phi$-$\psi$) again showing the inconsistency of the Ramachandran plot ($\phi$-$\psi$), see **Figure S2(e)** of the **Supplementary Materials**.

It can be seen from the conventional Ramachandran plot ($\phi$-$\psi$), as expected data points are most highly clustered around $\phi$ = -60° and $\psi$ = -50° [63] for lycosin-05 peptide, the global energy minimum structure, see **Figure S2(e)** in **Supplementary Materials S2**. Conversely, the least amount of clustering of data points on the conventional Ramachandran plot ($\phi$-$\psi$) occurs for the lycosin-02 peptide that possess the least energetically stable structure, see **Figure S2(b)** in **Supplementary Materials S2**. The Ramachandran plot ($\phi$-$\psi$) for each of the structures exhibit data points in the upper left region characteristic of the region of stability of the beta-sheet[64], the lycosin-02 peptide has four such data points. We suggest data points in the region of stability of the beta-sheet indicates the presence of a degree of beta-sheet *character* in these five α-helix structured peptides. The greater number of points indicating beta-sheet character present for the lycosin-02 peptide structure may explain the greatest energetic instability of the lycosin-02 peptide.

These trends towards the least and most clustering around $\phi$ = -60° and $\psi$ = -50° for the least (lyosin-02) and most (lyosin-05) energetically stable structures are also reflected in clustering of data points on the backbone QTAIM *interpreted* Ramachandran plot ($\beta_\phi$,$\beta_\psi$) in the region $\beta_\phi$ = -60° and $\beta_\psi$ = -50°, see **Figure 3(a)** and **Figure 3(e)** respectively. The definitions for the $\beta_\phi$ and $\beta_\psi$ that comprise the QTAIM *interpreted* Ramachandran plot ($\beta_\phi$,$\beta_\psi$) are provided in sub-figures **(a)** and **(b)** respectively of **Scheme 4**.

In **Figure 3** we focus on the ellipticity ε and not the equivalent plots color mapped to the stiffness $\mathbb{S}$ which can be found in the **Supplementary Materials S4**, because the ellipticity ε better reflects the differences in the 'bond-torsion' central to the construction of the Ramachandran plot ($\phi$-$\psi$). The relationship between the data points on the conventional Ramachandran plot ($\phi$-$\psi$) and the most preferred response ($\beta_\phi$,$\beta_\psi$) however, is non-linear, this can be seen by comparing each of the sub-figures of **Figure 3** with the corresponding sub-figures of **Supplementary Materials S2**.

The symmetry of the distribution of data points on both the most preferred response ($\beta_{\phi}$,$\beta_{\psi}$) and the least preferred response ($\beta^{*}_{\phi}$,$\beta^{*}_{\psi}$) are higher than that of the Ramachandran plot ($\phi$-$\psi$). This is explained by the shorter and folded range of values; ($\beta_{\phi}$,$\beta_{\psi}$) and ($\beta^{*}_{\phi}$,$\beta^{*}_{\psi}$) are both defined in the range, -90.0º ≤ ($\beta_{\phi}$,$\beta_{\psi}$),($\beta^{*}_{\phi}$,$\beta^{*}_{\psi}$) ≤ +90.0º, in contrast with the Ramachandran angles $\phi$ and $\psi$ that are defined in the range, -180.0º ≤ $\phi$,$\psi$ ≤ +180.0º. There is two-fold symmetry for (|$\beta^{*}_{\phi}$|,|$\beta^{*}_{\psi}$|) evident by the presence of two approximate mirror planes at $\beta^{*}_{\phi}$ = 0.0º and $\beta^{*}_{\psi}$ = 0.0º, see **Figure 4(a)**. In contrast, (|$\beta_{\phi}$|,|$\beta_{\psi}$|) only contains one mirror plane at $\beta_{\phi}$ = 0.0º, see **Figure 4(b)**. The clustering for (|$\beta^{*}_{\phi}$|,|$\beta^{*}_{\psi}$|) is compressed into to the corner regions 60.0º ≤ (|$\beta^{*}_{\phi}$|,|$\beta^{*}_{\psi}$|) ≤ 90.0º reflecting the stronger tendency for the bond-path framework to be more detached from the peptide backbone compared with (|$\beta_{\phi}$|,|$\beta_{\psi}$|). The higher degree of symmetry of (|$\beta^{*}_{\phi}$|,|$\beta^{*}_{\psi}$|) reflects the underlying higher symmetry properties of the constituent $\underline{\mathbf{e}}_1$ eigenvector, see equation **(5)**, compared with the $\underline{\mathbf{e}}_2$ eigenvector that comprises (|$\beta_{\phi}$|,|$\beta_{\psi}$|), see equation **(4)**. Previously, the $\underline{\mathbf{e}}_1$ eigenvector has been found to be more symmetrical from analysis of the basin path set area subjected to a torsion α[42] as compared with the $\underline{\mathbf{e}}_2$ eigenvector, see the theory section 2.2.

For the backbone shared-shell *BCP*s the regions occupied by (|$\beta^{*}_{\phi}$|,|$\beta^{*}_{\psi}$|) and (|$\beta_{\phi}$|,|$\beta_{\psi}$|) are complementary as was the case for the closed-shell *BCP*s, see **Figure 2**, compare **Figure 4(a)** and **Figure** 4**(b)**. The exception to this are indicated by the dark blue, brown or black squares, inside 60.0º ≤ (|$\beta_{\phi}$|,|$\beta_{\psi}$|) ≤ 90.0º, see **Figure 4(c)**. Comparison with the dark blue squares on the conventional Ramachandran plot ($\phi$-$\psi$) indicates a correspondence with the regions of beta-sheet stability close to the 'allowed' regions, the brown and black square correspond to the 'allowed' and 'unlikely' regions respectively, see **Supplementary Materials S2**.

Higher values of the ellipticity ε tend to correspond to lower values preferred response |$\beta_{\psi}$|, i.e. an increased degree of alignment of the bond-path framework with the peptide backbone as determined by values of $\beta_{\psi}$ closer to $\beta_{\psi}$ = 0.0º, see the red data points in **Figure 4(d)**. Conversely, no such correlation present between the ellipticity ε and the preferred response $\beta_{\phi}$. Higher values of the ellipticity ε also correspond to the 'favored' regions of the Ramachandran plot, consistent with higher double bond character, this can be seen by a process of elimination; none of the red data points are highlighted by squares corresponding to data points falling in the 'unlikely' and 'allowed' regions respectively.

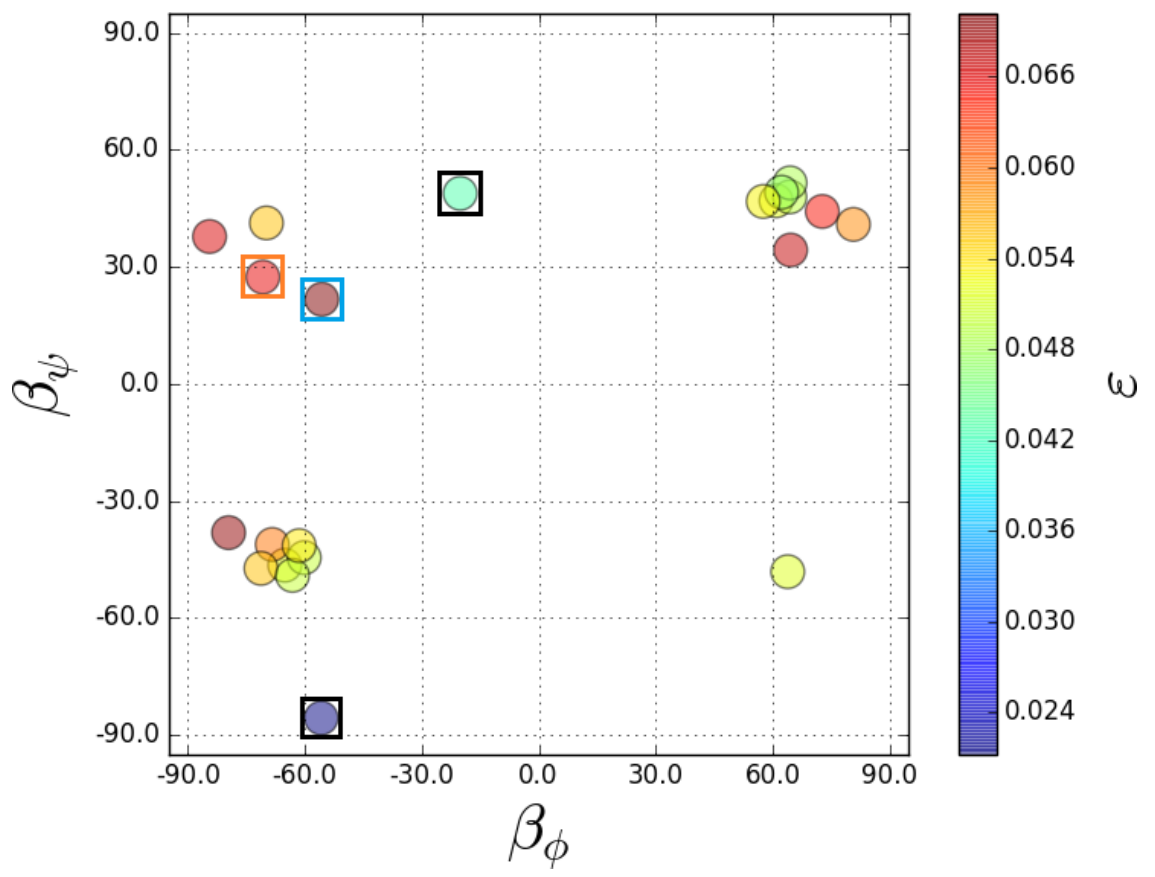

**(a)**

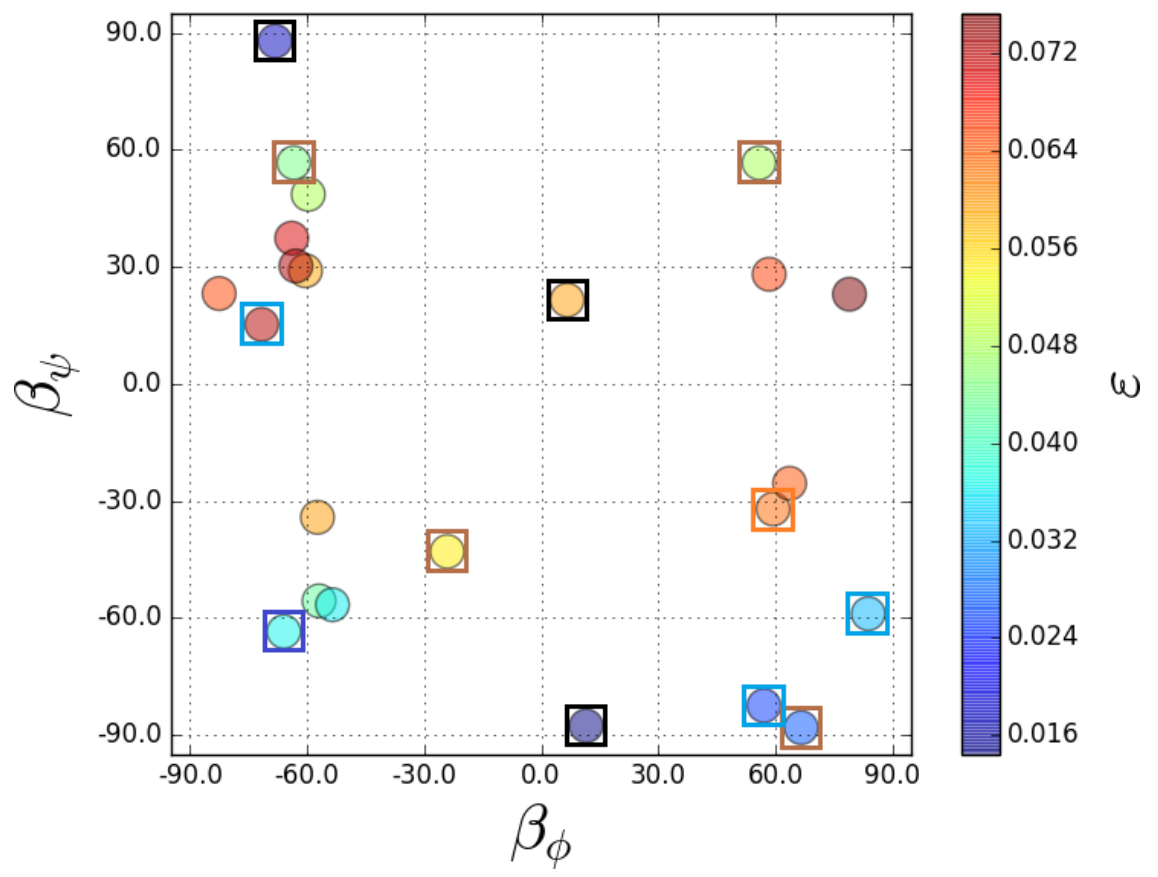

**(b)**

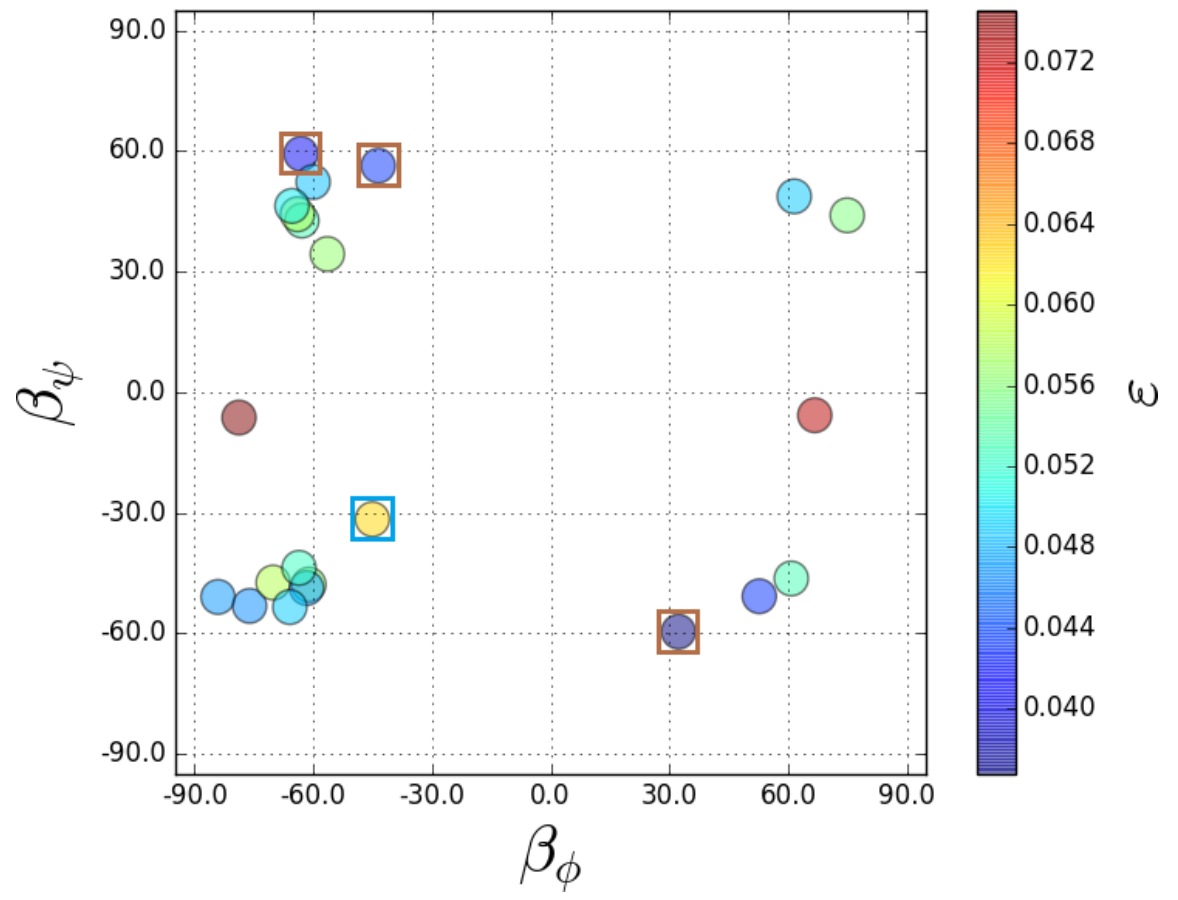

**(c)**

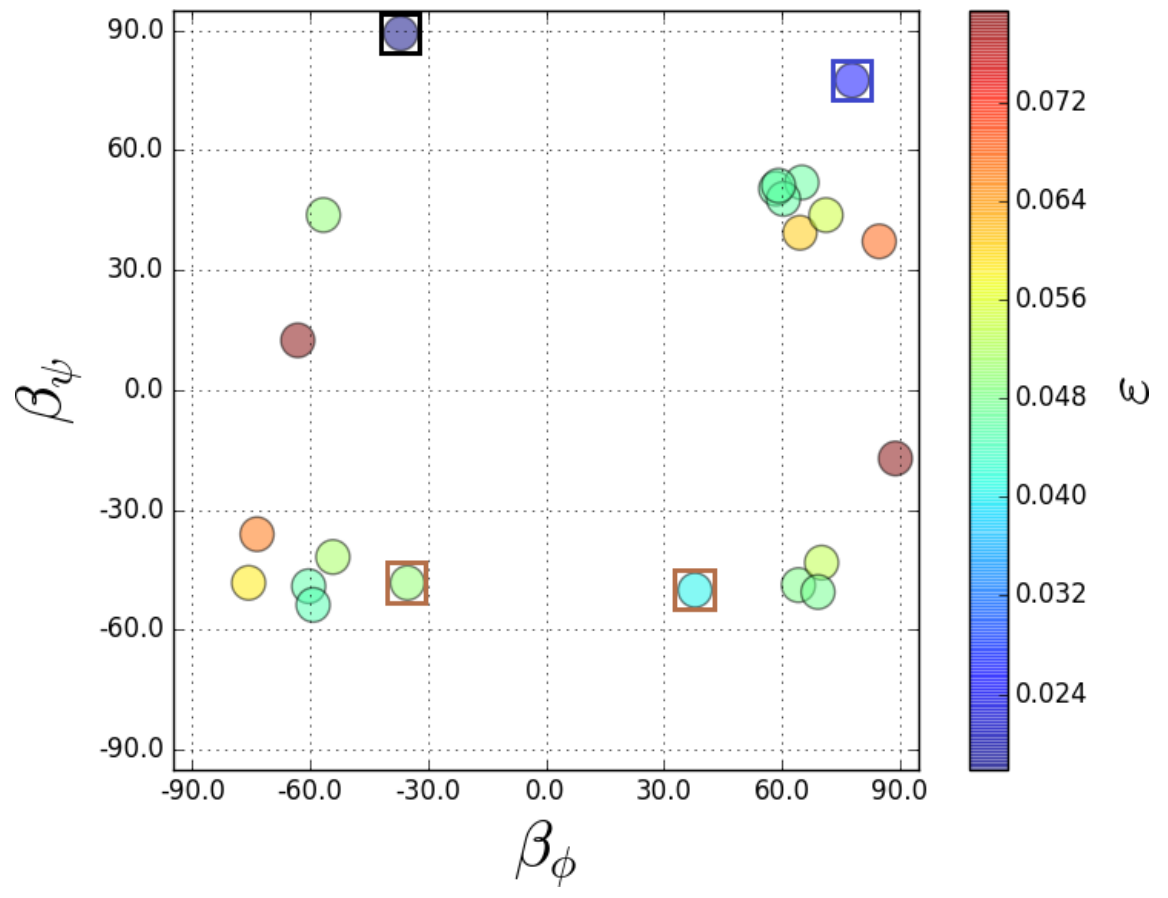

**(d)**

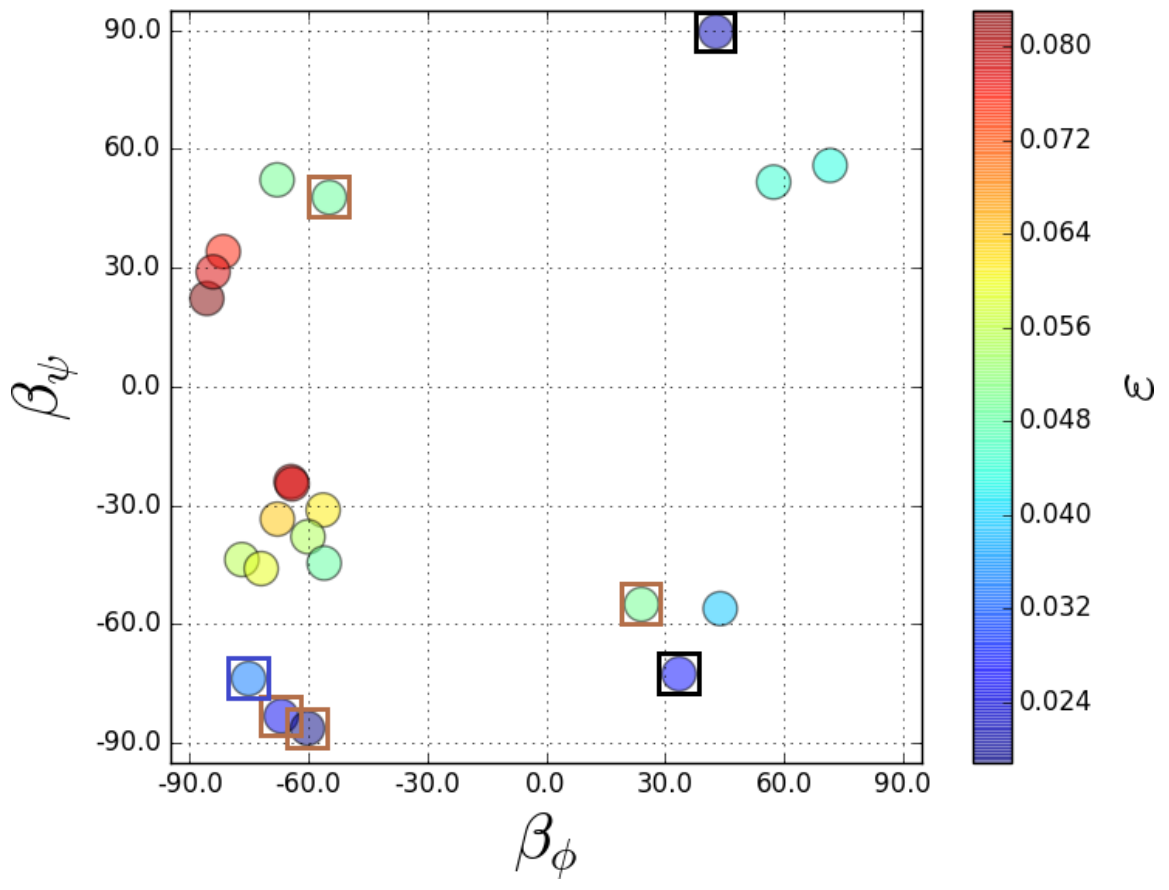

**(e)**

**Figure 3.** Plots of the response $\beta_\phi$ versus $\beta_\psi$ with the distribution of ellipticity ε of the backbone shared-shell *BCP*s for the molecular graph of lycosin-01, lycosin-02, lycosin-03, lycosin-04 and lycosin-05 are shown in sub-figures **(a)-(e).** The black, brown and dark blue square indicate the 'unlikely', 'allowed' and regions close to beta-sheet character from the conventional Ramachandran plot ($\phi$-$\psi$) respectively. The light blue square corresponds to points in the beta-sheet region on ($\phi$-$\psi$) shown in sub-figure **(a)-(c).** The orange squares indicate the amino acids residues corresponding to left-handed α-helices on ($\phi$-$\psi$) shown in sub-figures **(a)** and **(b)**. The definitions of $\beta_\phi$ and $\beta_\psi$ are given in **Scheme 4(a-b)** respectively. See **Supplementary Materials S2** and the figure caption of **Figure 2** for further details.

Therefore, to enable a clear understanding of the preferred response plots ($\beta_\phi$,$\beta_\psi$) we have indicated on the QTAIM plots for ($\beta_\phi$,$\beta_\psi$) the locations of the classic 'unlikely' and 'allowed' regions of the conventional Ramachandran plot ($\phi$-$\psi$) by black and brown squares respectively, compare **Figure 4(c)** and **Figure 4(e)**. The dark blue squares correspond to the 'stray' unexpected points that are found in the 60.0º ≤ ($|\beta_\phi|$,$|\beta_\psi|$) ≤ 90.0º region which have also been added onto the conventional Ramachandran plot ($\phi$-$\psi$), see the **Supplementary Materials S2**. The dark blue squares occur on the border of the beta-sheet character and 'allowed' regions of the conventional Ramachandran plot ($\phi$-$\psi$).

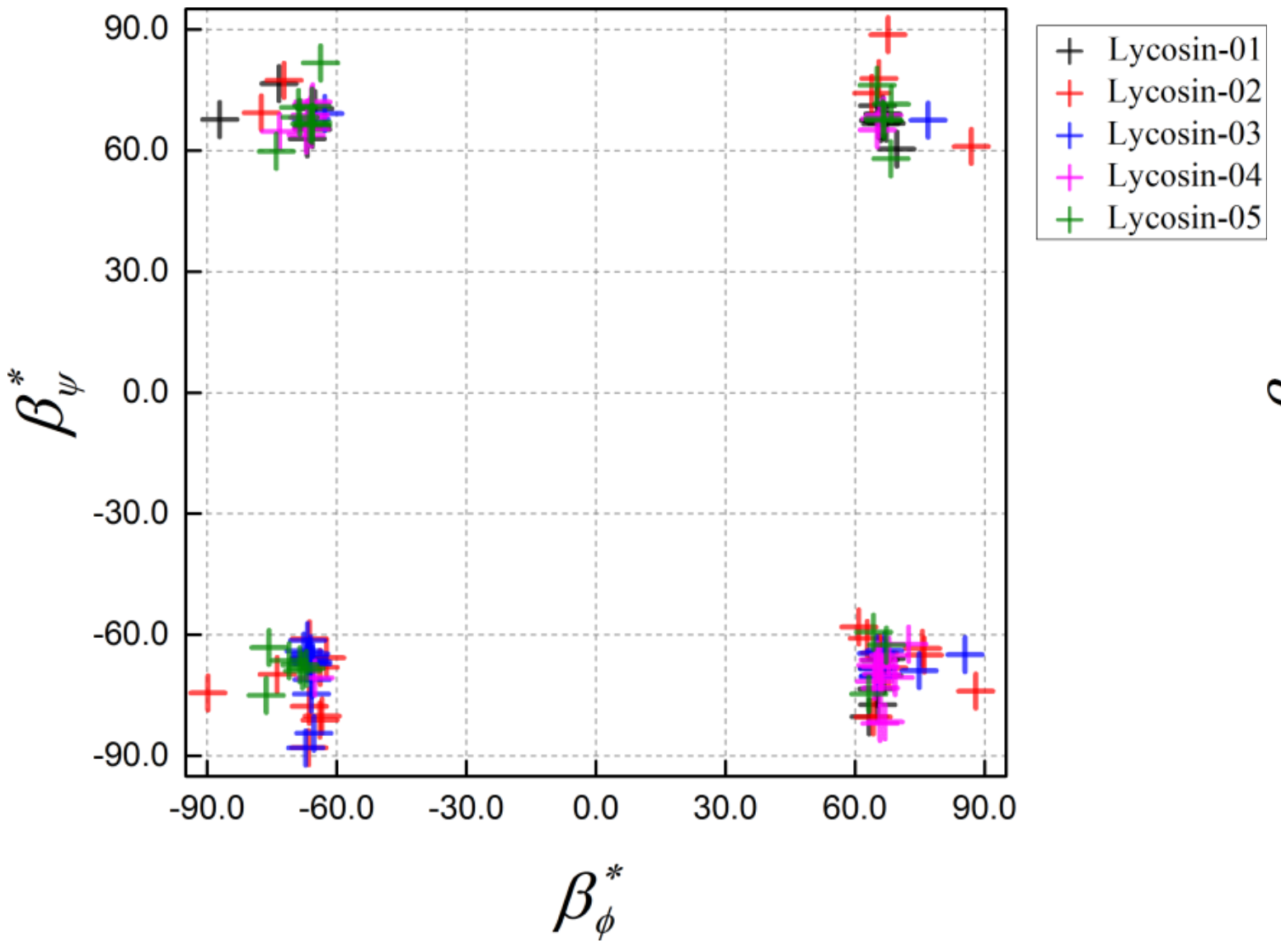

**(a)**

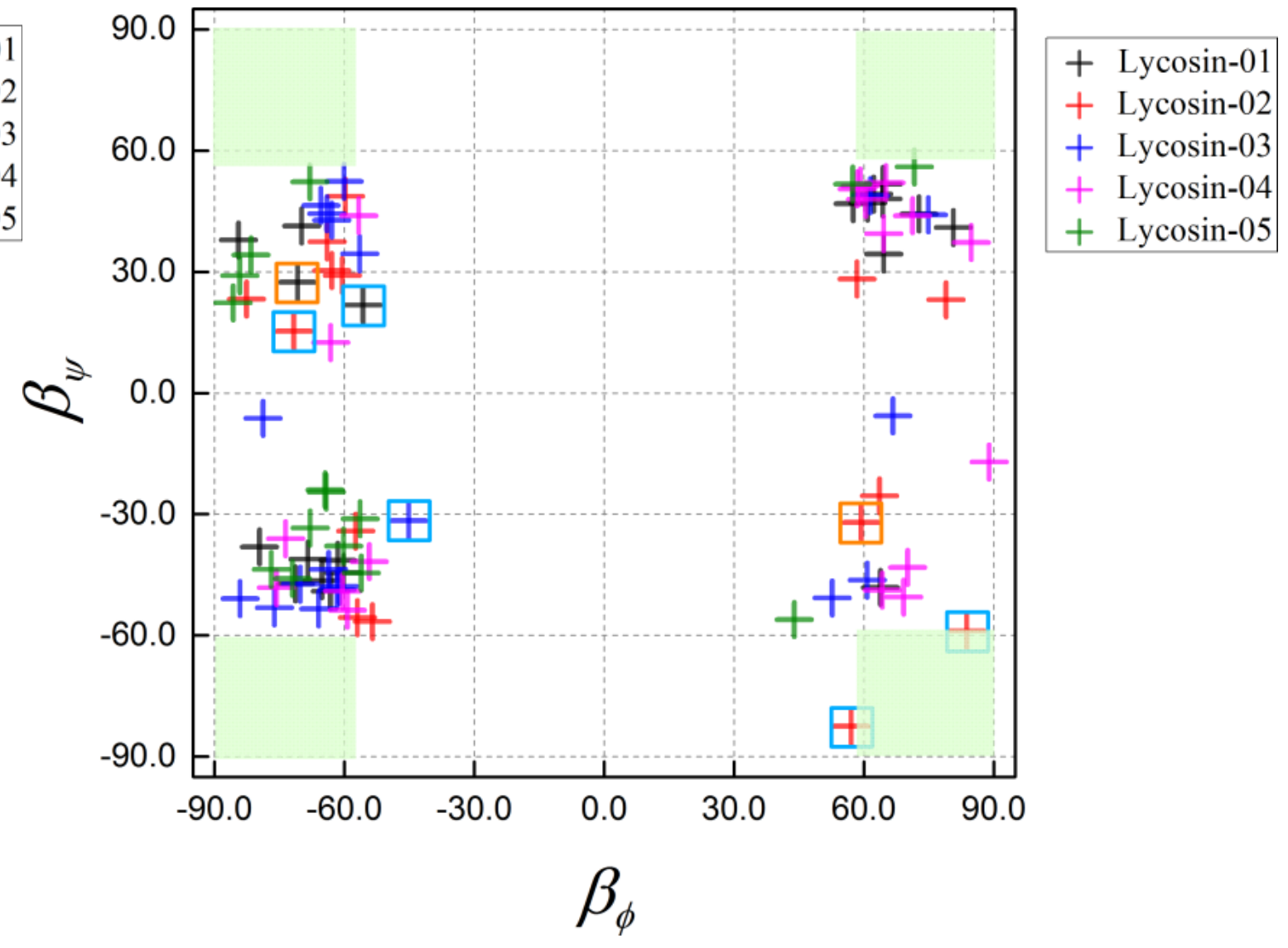

**(b)**

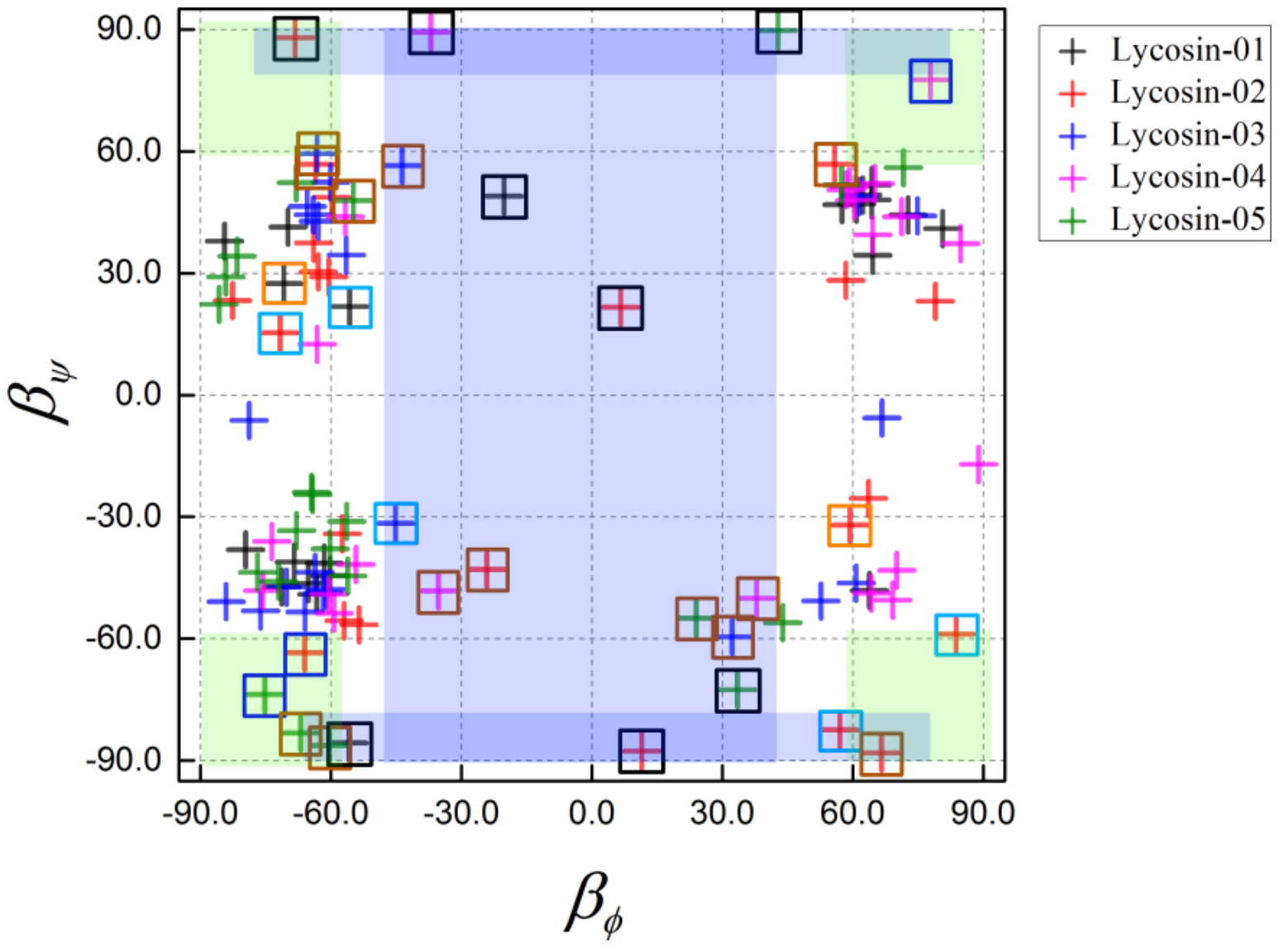

**(c)**

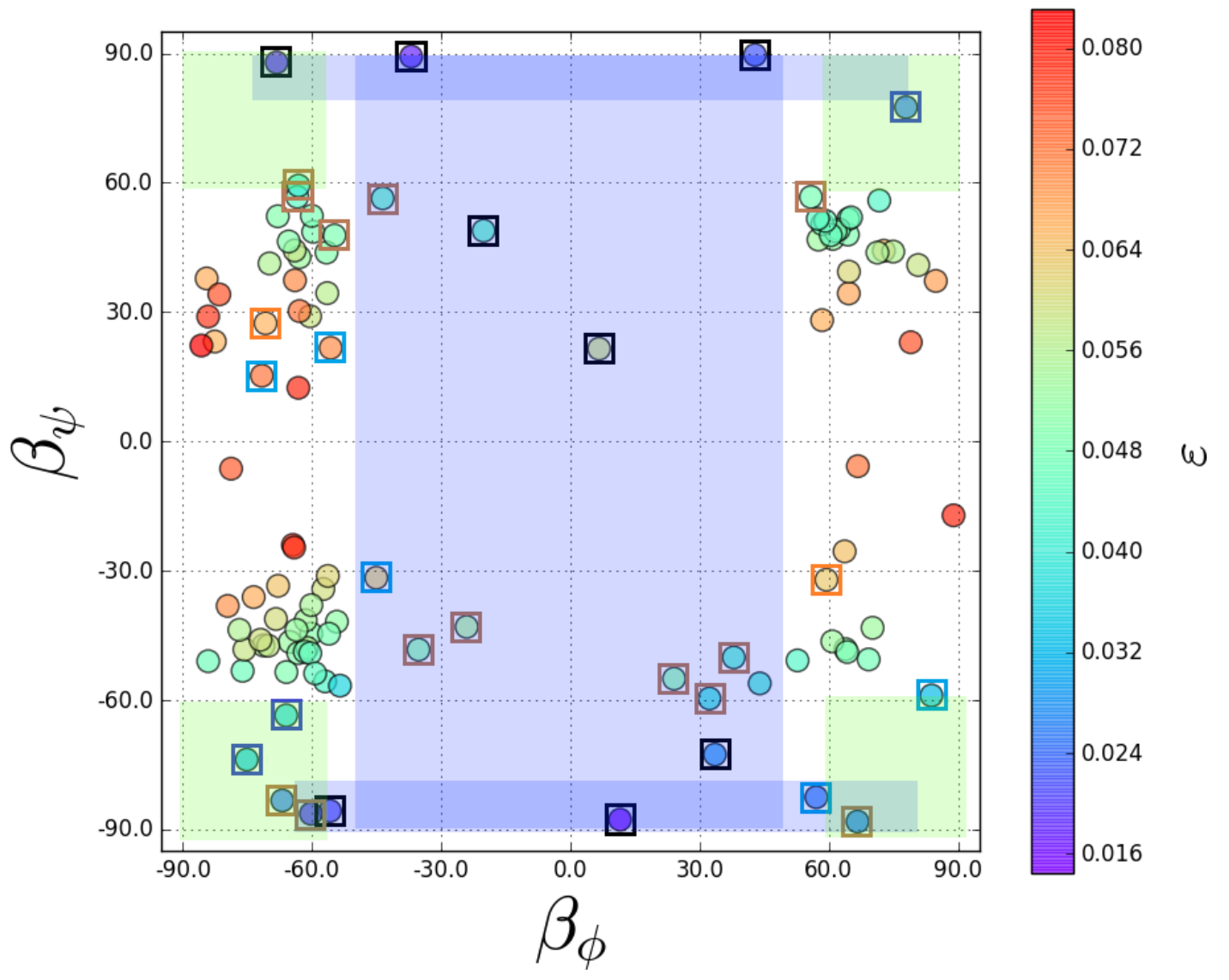

**(d)**

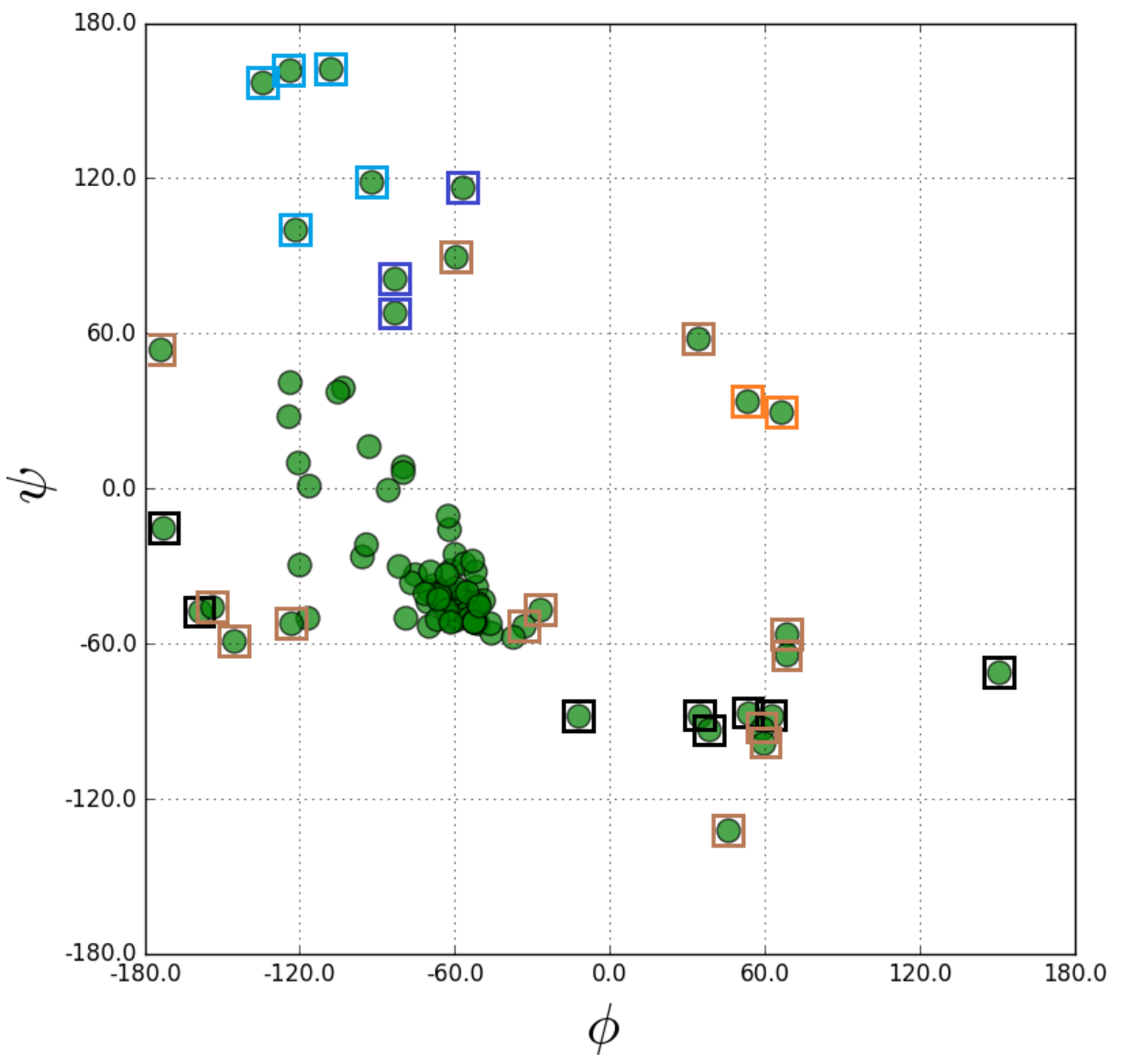

**(e)**

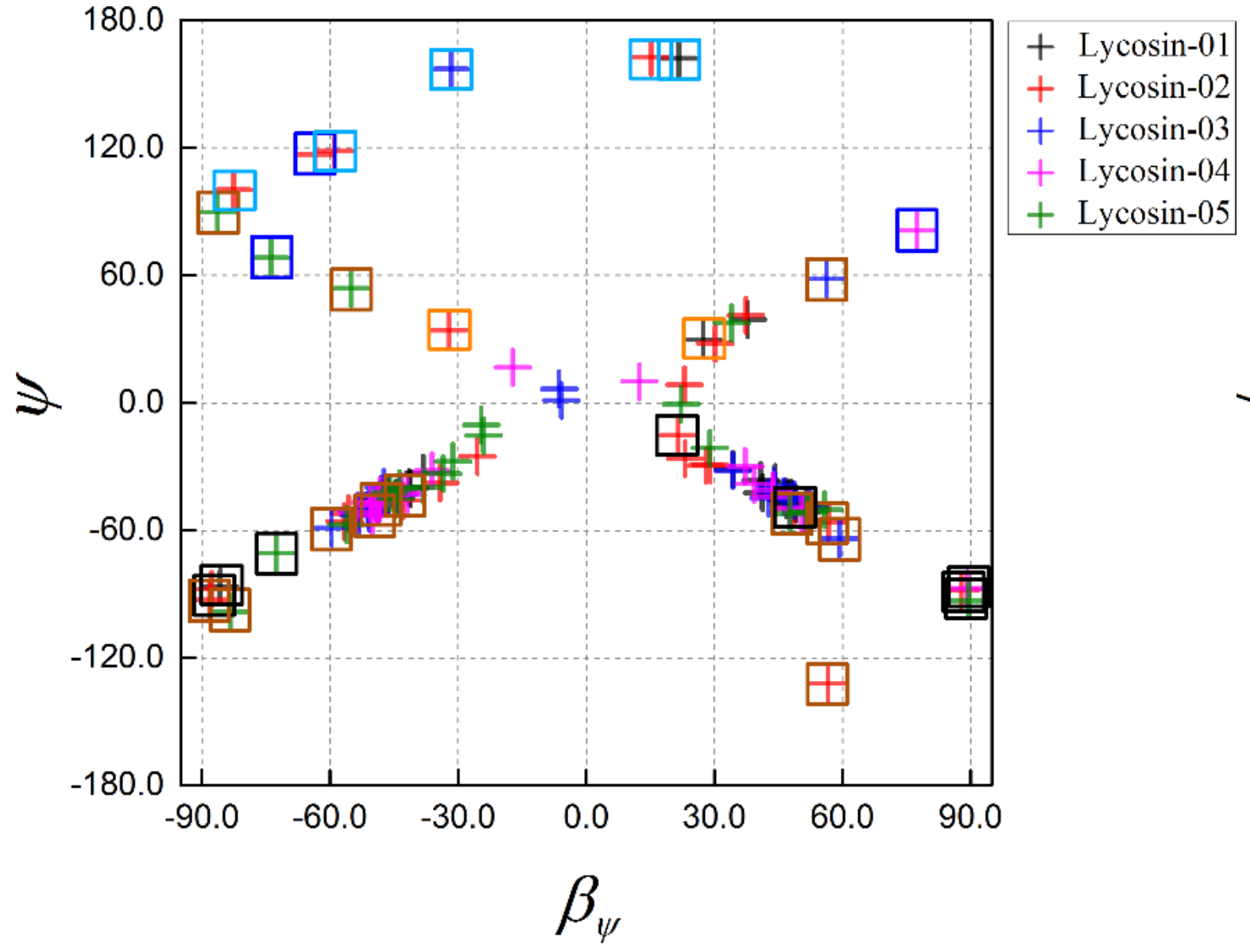

**(f)**

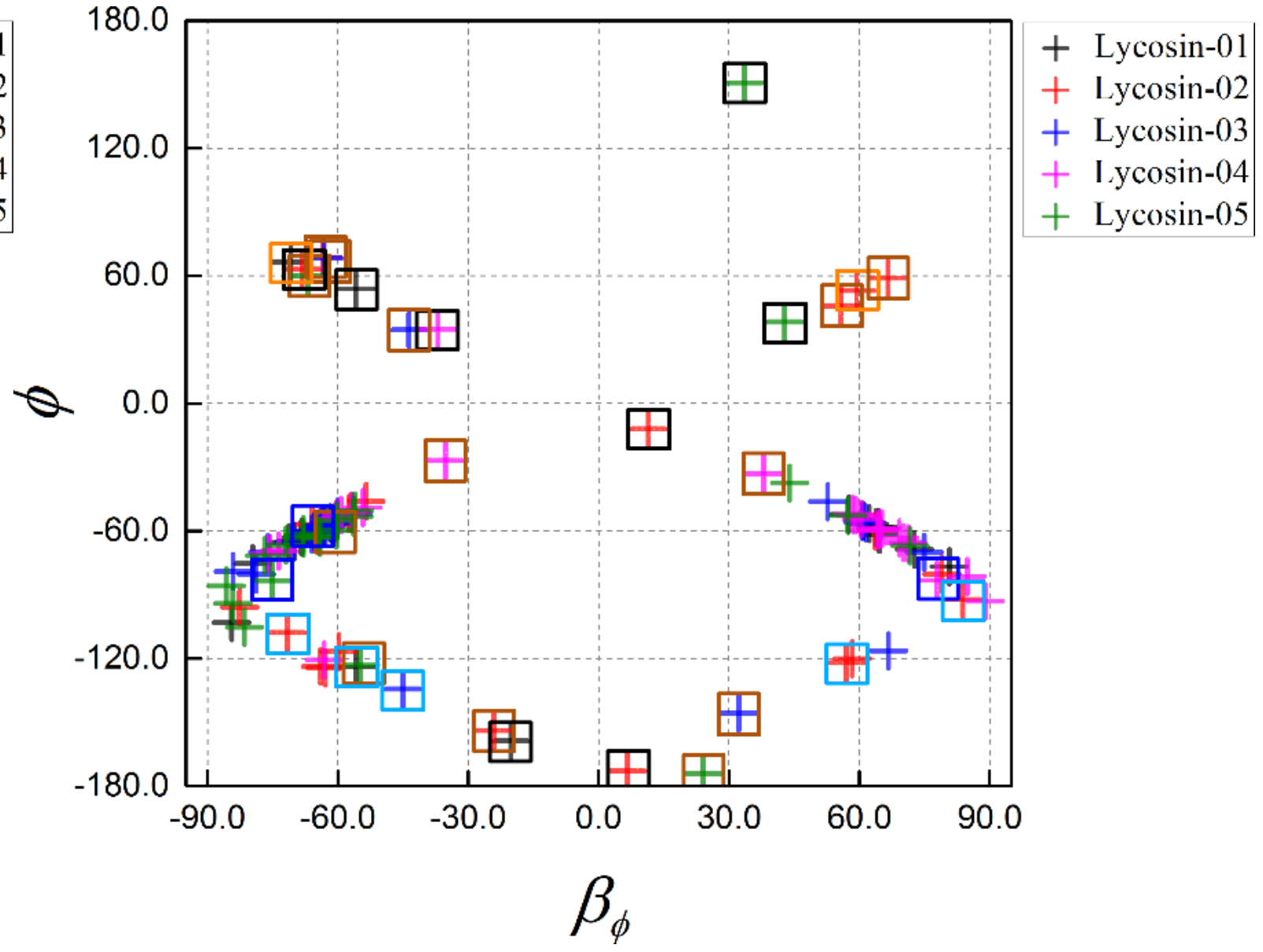

**(g)**

**Figure 4.** Plots of the all the response $\beta^{*}_{\phi}$ versus $\beta^{*}_{\psi}$ data with the distribution of the five lycosin peptides of this investigation are shown superimposed, in sub-figure **(a)**. The pale green square regions indicate the response ($\beta^{*}_{\phi}$,$\beta^{*}_{\psi}$) show the QTAIM 'allowed' region from QTAIM from **(a)**, the undecorated pluses, orange and pale blue squares correspond to the right handed α-helix, the left-handed α-helix and beta sheet regions of the Ramachandran plot ($\phi$-$\psi$) respectively are given sub-figure **(b)**, see also **Supplementary Materials S2.** The black, brown and dark blue square indicate the 'unlikely', 'allowed' and regions close to beta-sheet character from the conventional Ramachandran plot ($\phi$-$\psi$) respectively are added to sub-figure **(c)**, the approximate QTAIM 'unlikely' regions are indicated by the with blue rectangular regions, are given in sub-figure **(c)**. The plot of the responses ($\beta_{\phi}$,$\beta_{\psi}$) with the distribution of the ellipticity ε is given in sub-figure **(d)**. The raw distribution of the data for the five lycosin peptides from the Ramachandran plot ($\phi$-$\psi$) with the 'favored', 'allowed' and 'unlikely' regions omitted is provided for comparison in sub-figure **(e)**, the individual plots provided in **Supplementary Materials S2.** The plot of the distribution of $\psi$ from the Ramachandran plot against the response $\beta_{\psi}$ is given in sub-figure **(f)**. The corresponding plot for $\phi$ against $\beta_{\phi}$ is given in sub-figure **(g)**.

A backbone QTAIM interpreted Ramachandran plot ($\beta_{\phi}$,$\beta_{\psi}$) is created first from a superposition of the results for the five lycosin peptide conformers of the least preferred response ($\beta^{*}_{\phi}$,$\beta^{*}_{\psi}$) and is presented in **Figure 4(a)**, notice the highly localized distribution of data points. This localized region; 60.0º ≤ (|$\beta_{\phi}$|,|$\beta_{\psi}$|) ≤ 90.0º, occupied by ($\beta^{*}_{\phi}$,$\beta^{*}_{\psi}$) defines the 'allowed' regions on the QTAIM interpreted Ramachandran plot ($\beta_{\phi}$,$\beta_{\psi}$) and is indicated by the pale blue squares along with the data from the α- and beta-sheet regions as determined by the Ramachandran plot ($\phi$-$\psi$), see **Figure 4(b)**. Analogously, we also map the QTAIM 'unlikely' regions using ($\beta^{*}_{\phi}$,$\beta^{*}_{\psi}$) from the H--O/H---O *BCP*s and H---H *BCP*s, see **Figure 2(b)** with the blue rectangular regions for the five lycosin conformers, see **Figure 4(c)**. Comparison between the five lycosin conformers is provided by using the ellipticity ε, see **Figure 4(d)**.

The response $\beta_{\psi}$ is correlated with the Ramachandran angle $\psi$ as is the response $\beta_{\phi}$ with the corresponding Ramachandran angle $\phi$ indicated by the 'folded' linear relationships, see **Figure 4(f)** and **Figure 4(g)** respectively. Notice because the of the definition of the range of values, 60.0º ≤ (|$\beta_{\phi}$|,|$\beta_{\psi}$|) ≤ 90.0º, the plots $\beta_{\psi}$ vs. $\psi$ and $\beta_{\phi}$ vs. $\phi$ is 'folded' inwards. These plots therefore suggest that the responses $\beta_{\phi}$,$\beta_{\psi}$ intrinsically contain secondary structure information

## Conclusions

In this investigation we have introduced a QTAIM interpretation of the Ramachandran plot ($\phi$-$\psi$) for the closed-shell H--O/H---O *BCP*s and H---H *BCP*s as well as the backbone shared-shell *BCP*s, where *BCP* denotes bond critical point. Consistency was found between the QTAIM interpreted Ramachandran plot ($\beta_{\phi}$,$\beta_{\psi}$) and the corresponding Ramachandran plot ($\phi$-$\psi$) by a point-by-point mapping between the two representations. The relationship between the response plot ($\beta_{\phi}$,$\beta_{\psi}$) and the conventional Ramachandran plot ($\phi$-$\psi$) is highly non-linear. This shows that this new approach to examine the topology of peptides is not a trivial coordinate transformation, or equivalent, but contains fundamentally new information from the $\underline{\mathbf{e}}_{1}$ and $\underline{\mathbf{e}}_{2}$ eigenvectors of the total electronic charge density $\rho(\mathbf{r_b})$. Therefore, no information is lost from the conventional Ramachandran plot ($\phi$-$\psi$) and we have added information beyond nuclear geometries from the inherent tendencies stored in the $\underline{\mathbf{e}}_{1}$ and $\underline{\mathbf{e}}_{2}$ eigenvectors for the total electronic charge density $\rho(\mathbf{r_b})$ to accumulate in the least and most preferred directions respectively.

The scalar and vector derived properties of a series of five lycosin peptides from the QTAIM revealed the importance of the H--O/H---O *BCP*s and H---H *BCP*s in peptide folding preferences. The non-preferred orientations of the H--O/H---O *BCP*s, given by the response ($\beta^{*}_{\phi}$,$\beta^{*}_{\psi}$), coincided with the 'unlikely' regions of the conventional Ramachandran plot ($\phi$-$\psi$). This shows that the response ($\beta^{*}_{\phi}$,$\beta^{*}_{\psi}$) provides direct consideration of the least preferred orientations of the H--O/H---O *BCP*s and H---H *BCP*s and hence the regions of avoided folding preferences on the conventional Ramachandran plot ($\phi$-$\psi$). The results for the H--O/H---O *BCP*s were found to follow physical intuition, in that higher values of the stiffness $\mathbb{S}$ corresponded to larger values of the most preferred response $\beta_{\psi}$ and hence indicated a greater degree of *detachment* of the {$\underline{\mathbf{e}}_1$, $\underline{\mathbf{e}}_2$, $\underline{\mathbf{e}}_3$} framework from the containing nuclear skeleton. The converse was found for the correlation between the stiffness $\mathbb{S}$ values of the H--O/H---O *BCP*s and the least preferred response $\beta^{*}_{\psi}$.

In addition a backbone QTAIM *interpreted* Ramachandran plot ($\beta_{\phi}$,$\beta_{\psi}$) for the backbone shared-shell *BCP*s was uncovered. This new approach was formulated to match as closely as possible the construction of the conventional Ramachandran plot ($\phi$-$\psi$) with the addition of being informed by the QTAIM vector derived properties in the form of the preferred response $\beta_{\phi}$ and $\beta_{\psi}$. Specifically, we found that lower values of the ellipticity ε of the backbone shared-shell *BCP*s correlated with greater degree of detachment of the {$\underline{\mathbf{e}}_1$, $\underline{\mathbf{e}}_2$, $\underline{\mathbf{e}}_3$} framework from the containing nuclear skeleton as measured by larger values of the most preferred response $\beta_{\psi}$. We see that the highest values of the ellipticity ε also tends to correspond to the 'favored' regions of the conventional Ramachandran plot ($\phi$-$\psi$), conversely the lowest values of the ellipticity ε of the backbone shared-shell *BCP*s correspond to the 'unlikely', 'allowed' and beta-sheet regions of the conventional Ramachandran plot ($\phi$-$\psi$).

Future applications could include polymers or molecular clusters where folding or torsion phenomena are particularly important, such applications include sugar saccharides or molecular rotary motors [65,66].

## Acknowledgements

The National Natural Science Foundation of China is gratefully acknowledged, project approval number: 21673071. The One Hundred Talents Foundation of Hunan Province and the aid program for the Science and Technology Innovative Research Team in Higher Educational Institutions of Hunan Province are also gratefully acknowledged for the support of S.J. and S.R.K. S.M. was supported by the Ministry of Education of Singapore.

# ELECTRONIC SUPPLEMENTARY INFORMATION

A QTAIM Interpretation of the Ramachandran Plot ($\phi$-$\psi$) for Peptides

**Roya Momen[1], Alireza Azizi[1], Lingling Wang [1], Ping Yang [1], Tianlv Xu[1], Steven R. Kirk[*1] Wenxuan Li[2], Sergei Manzhos[2] and Samantha Jenkins[*1]**

*[1]Key Laboratory of Chemical Biology and Traditional Chinese Medicine Research and Key Laboratory of Resource Fine-Processing and Advanced Materials of Hunan Province of MOE, College of Chemistry and Chemical Engineering, Hunan Normal University, Changsha, Hunan 410081, China*
*[2]Department of Mechanical Engineering, National University of Singapore, Block EA 07-08, 9 Engineering Drive 1, Singapore 117576*

*email: samanthajsuman@gmail.com
*email: steven.kirk@cantab.net

## Contents:

**1. Supplementary Materials S1.**

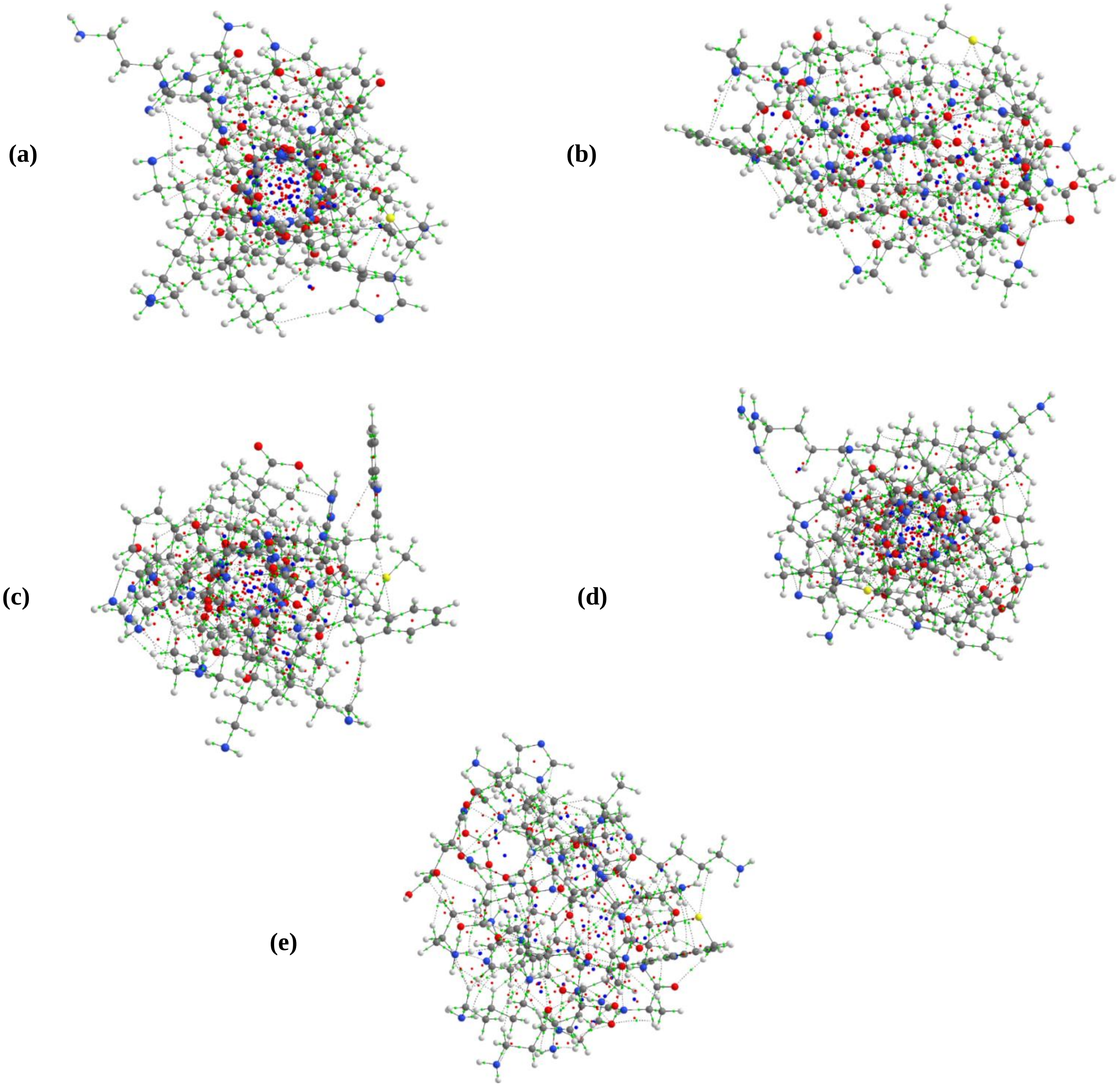

**Figure S1.** The molecular graphs of the five lycosin peptide structures, lycosin-01, lycosin-02, lycosin-03, lycosin-04 and lycosin-05 are presented in sub-figures **(a)-(e)** respectively. The green, small red and small blue spheres represent the bond critical points *BCPs*, ring critical points *RCPs* and cage critical points *CCPs*, respectively. For a summary of the QTAIM topology and relative energies ΔE see **Table 1**.

**2. Supplementary Materials S2.** The Ramachandran plots ($\phi$-$\psi$) for five conformers of the lycosin peptide with the amino acid sequence RKGWFKAMKSIAKFIAKEKLKEHL. The data points corresponding to the backbone in the lycosin peptide backbone are represented by the green circles. The magenta, blue and white regions correspond to the 'favored', 'allowed' and 'unlikely' regions respectively.

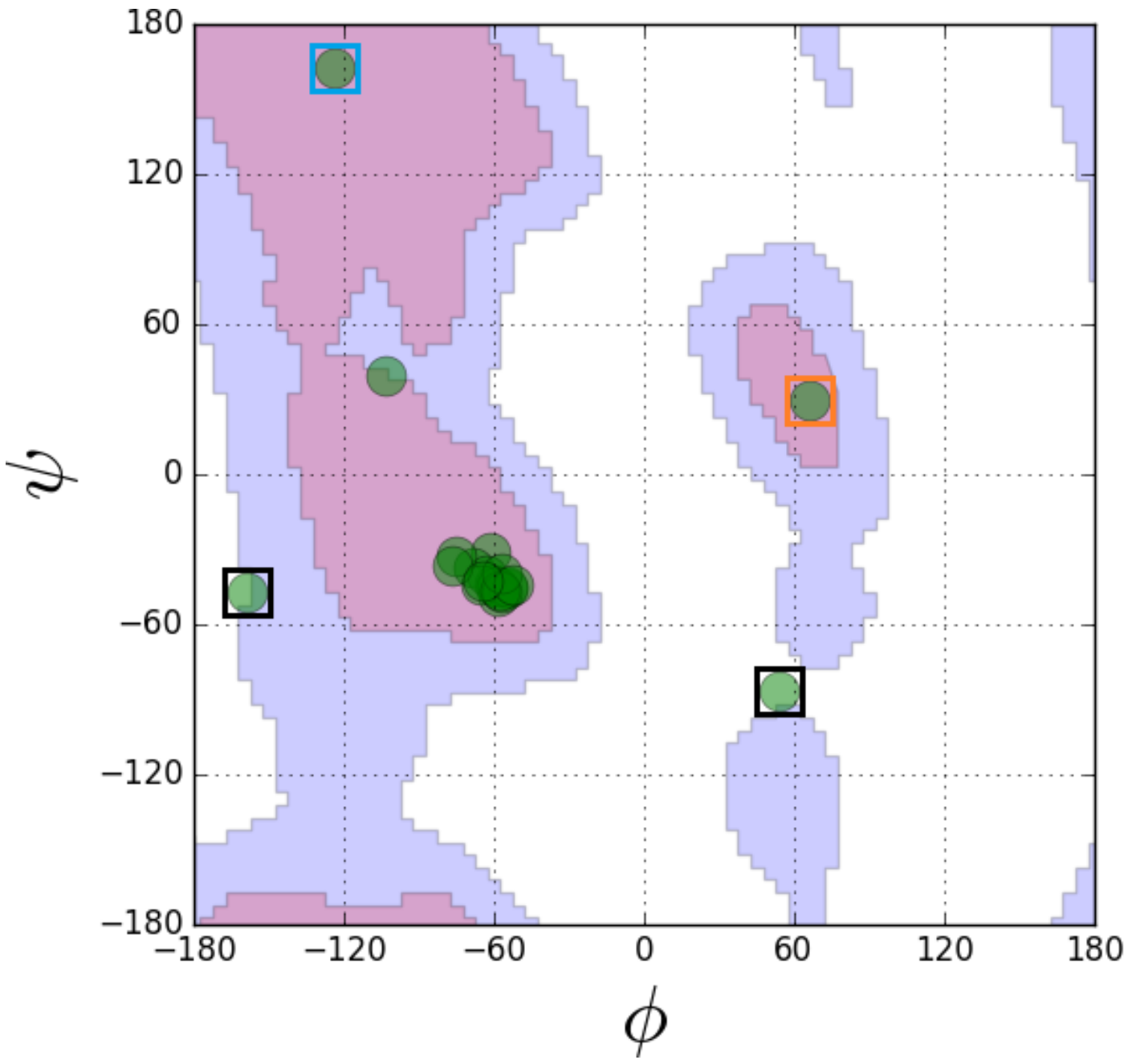

**(a)**

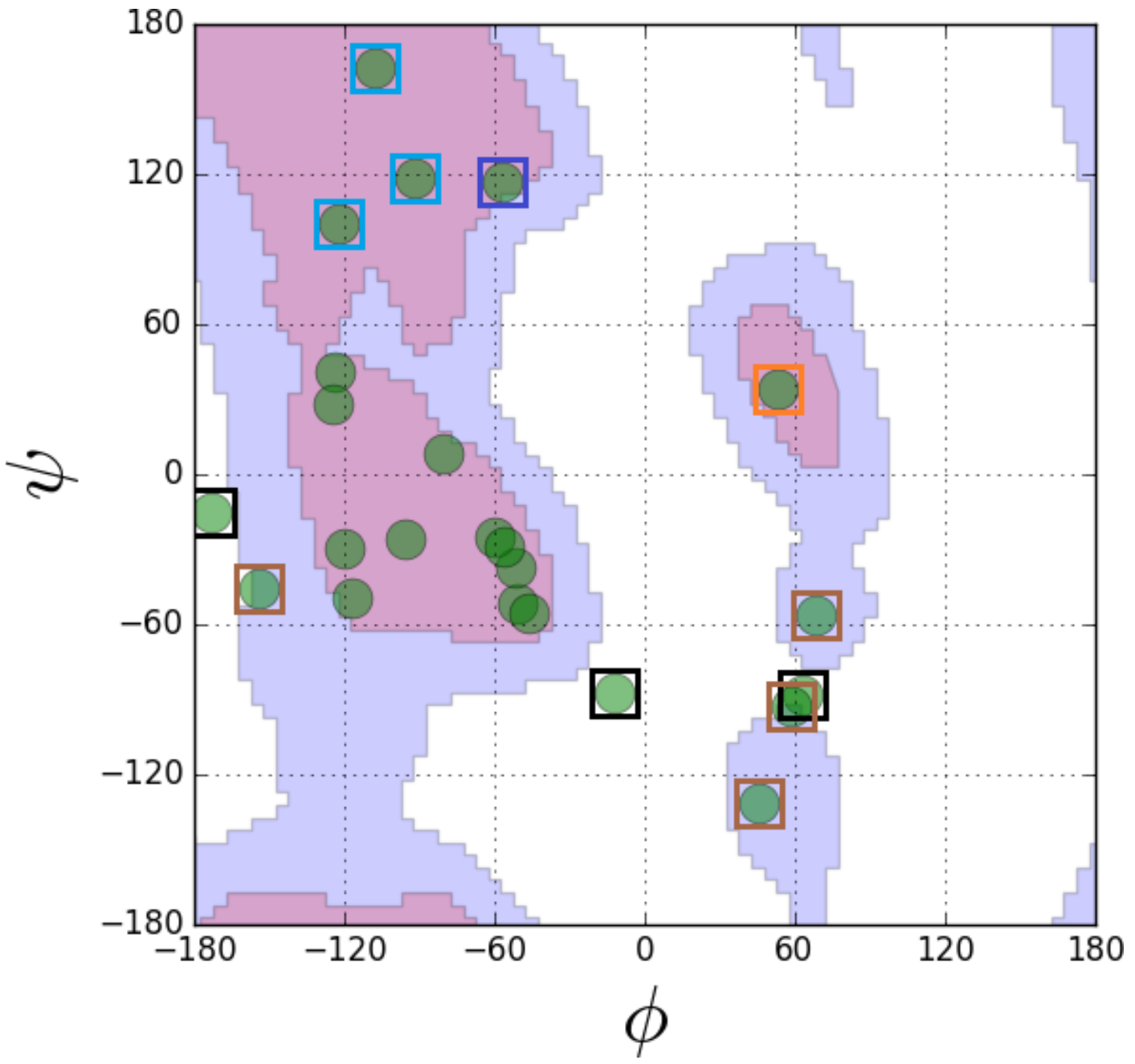

**(b)**

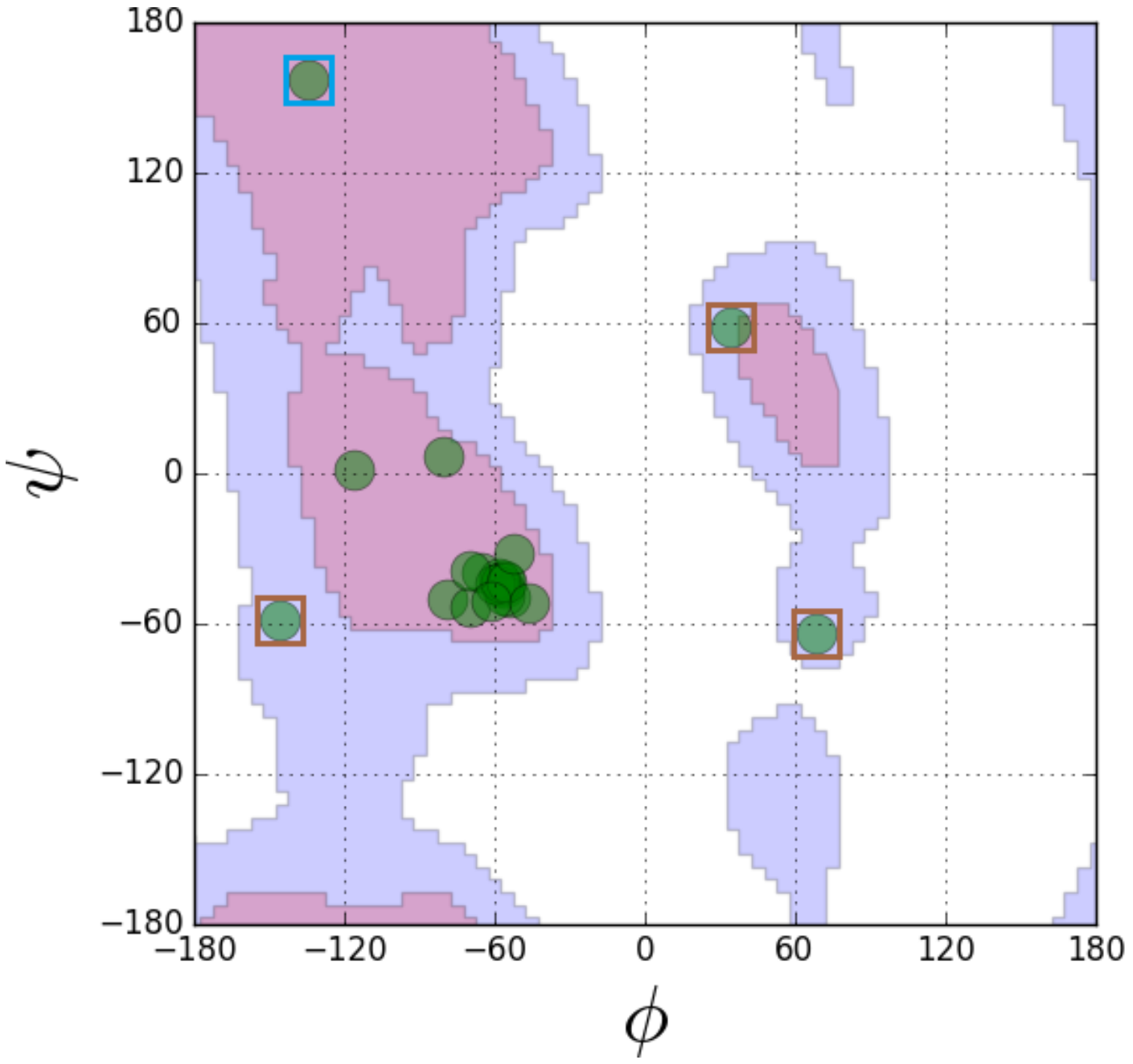

**(c)**

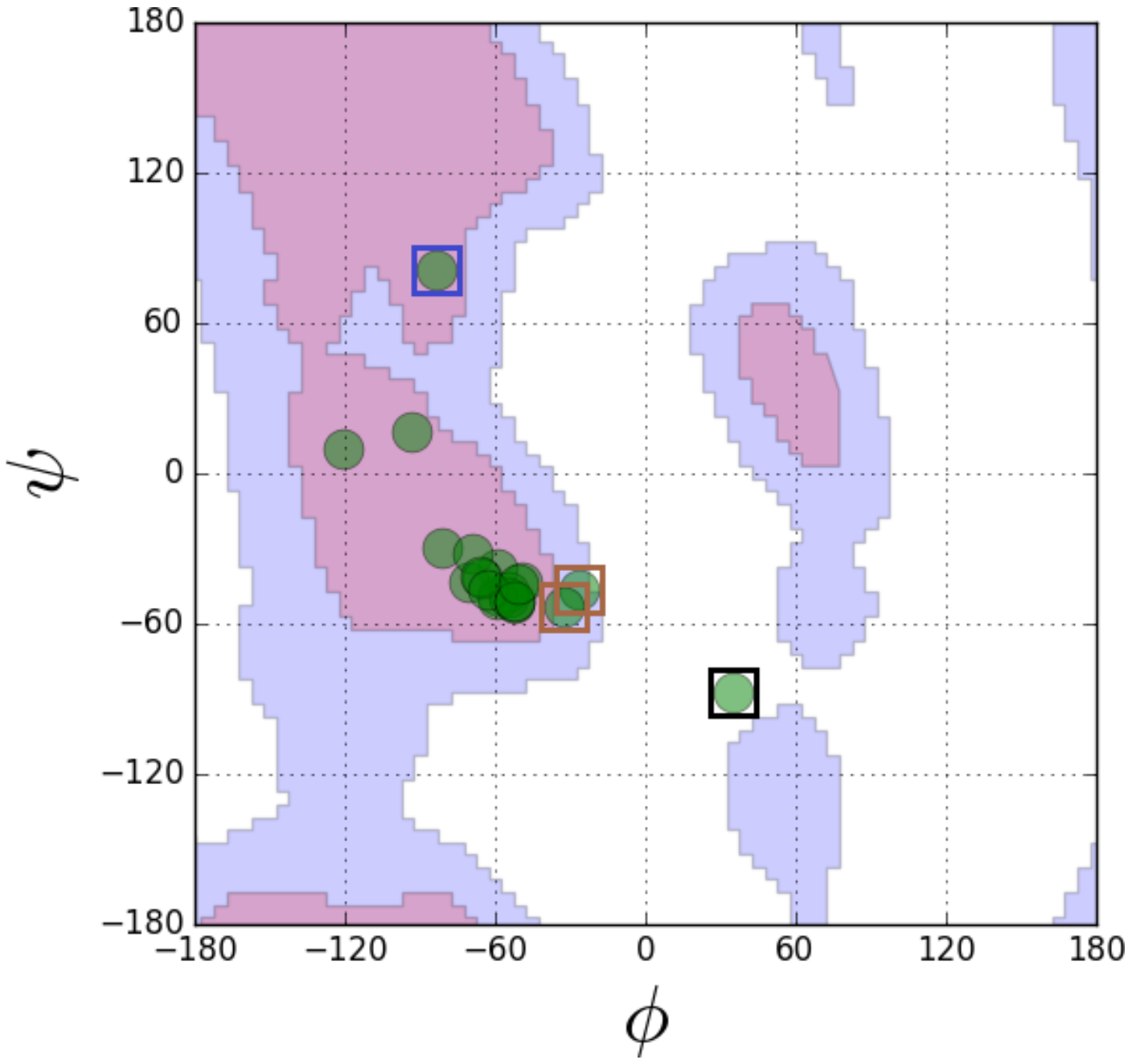

**(d)**

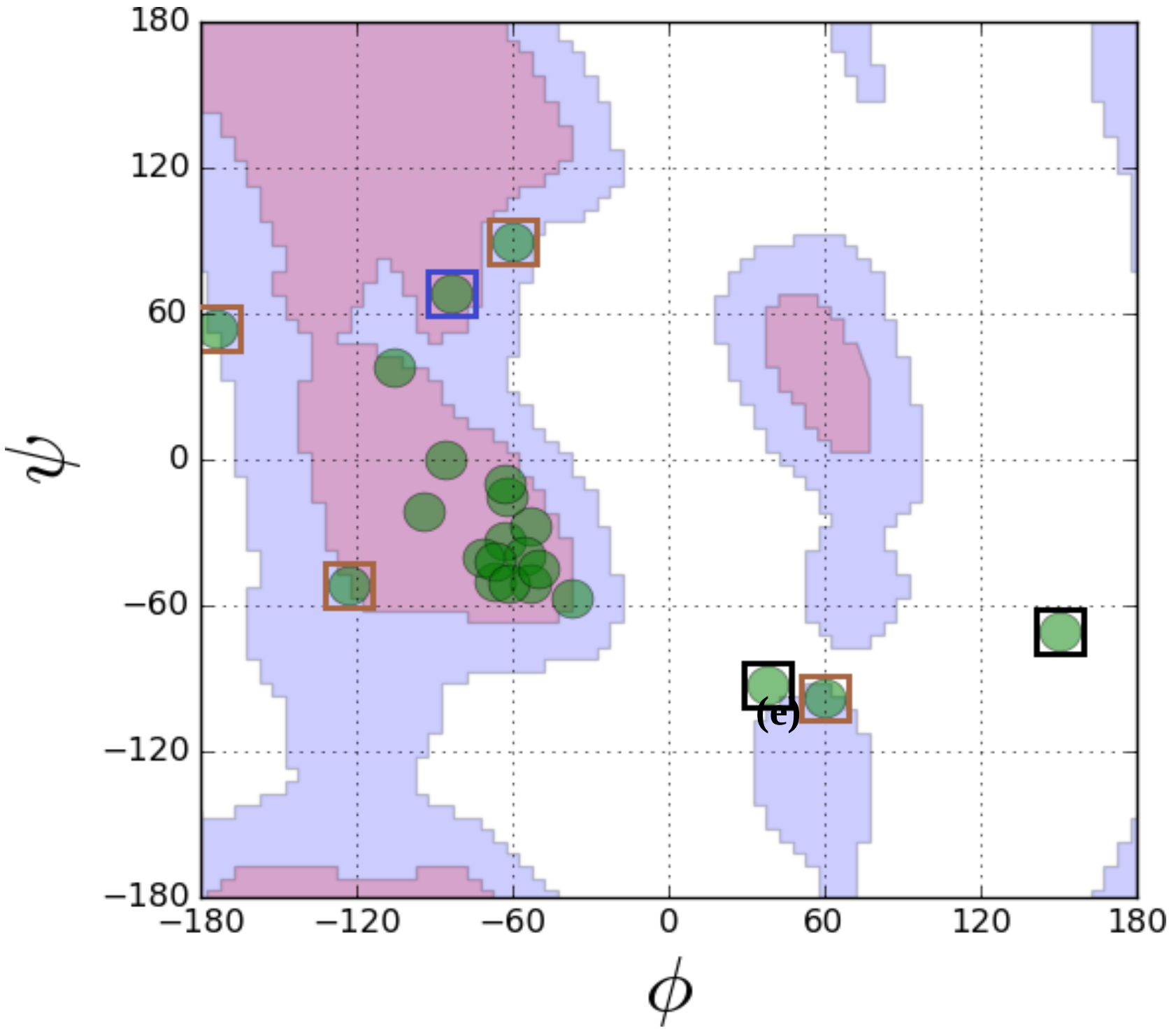

**(e)**

**Figure S2.** The Ramachandran ($\phi$-$\psi$) plots for lycosin-01, lycosin-02, lycosin-03, lycosin-04 and lycosin-05 peptide structures are presented in sub-figures **S2(a)-(e)** respectively. The 'unlikely' and 'allowed' regions correspond to the white and blue regions indicated where the data points are highlighted by black and brown squares respectively. The light blue squares in sub-figures **S2(a)**, **S2(b)** and **S2(c)** indicate beta-sheet character points, see **Figure 3(a)**, **Figure 3(b)** and **Figure 3(c)** in the main paper respectively, also see **Table S2**. The dark blue squares indicate beta-sheet character points close to the 'allowed' region. The orange squares in sub-figures **S2(a)** and **S2(b)** indicate the left-handed helix data points, see **Figure 3(a)** and **Figure 3(b)** in the main paper respectively.

**Table S2**. The mapping of the conventional Ramachandran plot ($\phi$-$\psi$) for the five lycosin peptide structures; lycosin-01, lycosin-02, lycosin-03, lycosin-04 and lycosin-05, to the QTAIM interpreted Ramachandran plot ($\beta_\phi$, $\beta_\psi$). The values ($\phi$, $\psi$) and ($\beta_\phi$, $\beta_\psi$) are angles that are specified in degrees (º). See the caption of **Figure S2** for an explanation of the Ramachandran plot ($\phi$-$\psi$) regions and the associated colored squares.

**Note this table will not be visible in its entirety if printed out on Letter format paper.**

| Peptide molecular graph | Conventional Ramachandran plot ($\phi$-$\psi$) regions | | | | | | | | QTAIM interpreted Ramachandran plot | |
|---|---|---|---|---|---|---|---|---|---|---|
| | 'unlikely' (white) | | 'allowed' (blue) | | beta sheet close to 'allowed' (dark blue square) | | beta sheet (light blue square) | | ($\beta_\phi$, $\beta_\psi$) | |
| | $\phi$ | $\psi$ | $\phi$ | $\psi$ | $\phi$ | $\psi$ | $\phi$ | $\psi$ | $\beta_\phi$ | $\beta_\psi$ |
| Lycosin-01 | 53.7235 | -86.9300 | --- | --- | --- | --- | --- | --- | -55.8170 | -85.6736 |
| | -158.9007 | -47.2171 | --- | --- | --- | --- | --- | --- | -20.1533 | 48.8773 |
| | --- | --- | --- | --- | --- | --- | -124.0719 | 162.0175 | -55.6606 | 21.7034 |
| Lycosin-02 | -172.9432 | -15.2848 | --- | --- | --- | --- | --- | --- | 6.6562 | 21.5151 |
| | -12.1767 | -87.7989 | --- | --- | --- | --- | --- | --- | 11.5159 | -87.6969 |
| | 62.9196 | -88.1855 | --- | --- | --- | --- | --- | --- | -68.2114 | 87.9098 |
| | --- | --- | 58.8300 | -92.8061 | --- | --- | --- | --- | 66.6437 | -88.1524 |
| | --- | --- | -153.9809 | -46.0455 | --- | --- | --- | --- | -24.0885 | -42.9929 |
| | --- | --- | 68.5818 | -56.4176 | --- | --- | --- | --- | -63.4202 | 56.7291 |
| | --- | --- | 45.6368 | -132.0922 | --- | --- | --- | --- | 55.8425 | 56.7012 |
| | --- | --- | --- | --- | --- | --- | -107.9072 | 162.5882 | -71.6414 | 15.2530 |
| | --- | --- | --- | --- | --- | --- | -92.3041 | 118.4916 | 83.8259 | -58.9082 |
| | --- | --- | --- | --- | --- | --- | -122.0578 | 100.1144 | 57.0794 | -82.4984 |
| | --- | --- | --- | --- | -57.0358 | 116.8138 | --- | --- | -65.9817 | -63.5166 |
| Lycosin-03 | --- | --- | 34.4555 | 58.2217 | --- | --- | --- | --- | -43.5887 | 56.3951 |
| | --- | --- | -145.7639 | -59.0927 | --- | --- | --- | --- | 32.2779 | -59.6150 |
| | --- | --- | 68.1933 | -64.0031 | --- | --- | --- | --- | -63.1287 | 59.3394 |
| | --- | --- | --- | --- | --- | --- | -134.4785 | 157.0556 | -45.0755 | -31.6090 |
| Lycosin-04 | 34.8592 | -87.7199 | --- | --- | --- | --- | --- | --- | -36.9709 | 89.2843 |
| | --- | --- | -26.9639 | -46.9602 | --- | --- | --- | --- | -35.2641 | -48.2717 |
| | --- | --- | -33.0279 | -53.2460 | --- | --- | --- | --- | 37.9319 | -50.0900 |
| | --- | --- | --- | --- | -83.3098 | 81.0172 | --- | --- | 77.9101 | 77.5184 |
| Lycosin-05 | 150.5167 | -70.8660 | --- | --- | --- | --- | --- | --- | 33.5537 | -72.5936 |
| | 38.1490 | -93.2386 | --- | --- | --- | --- | --- | --- | 42.8538 | 89.6388 |
| | --- | --- | 59.7033 | -98.5206 | --- | --- | --- | --- | -66.8995 | -83.2578 |
| | --- | --- | -59.7199 | 89.5797 | --- | --- | --- | --- | -60.2545 | -86.3395 |
| | --- | --- | -174.2406 | 53.7526 | --- | --- | --- | --- | 24.0662 | -55.0155 |
| | --- | --- | -123.2160 | -52.3249 | --- | --- | --- | --- | -54.7672 | 47.7957 |
| | --- | --- | --- | --- | -83.4680 | 68.1928 | --- | --- | -75.1358 | -73.7802 |

**3. Supplementary Materials S3.**

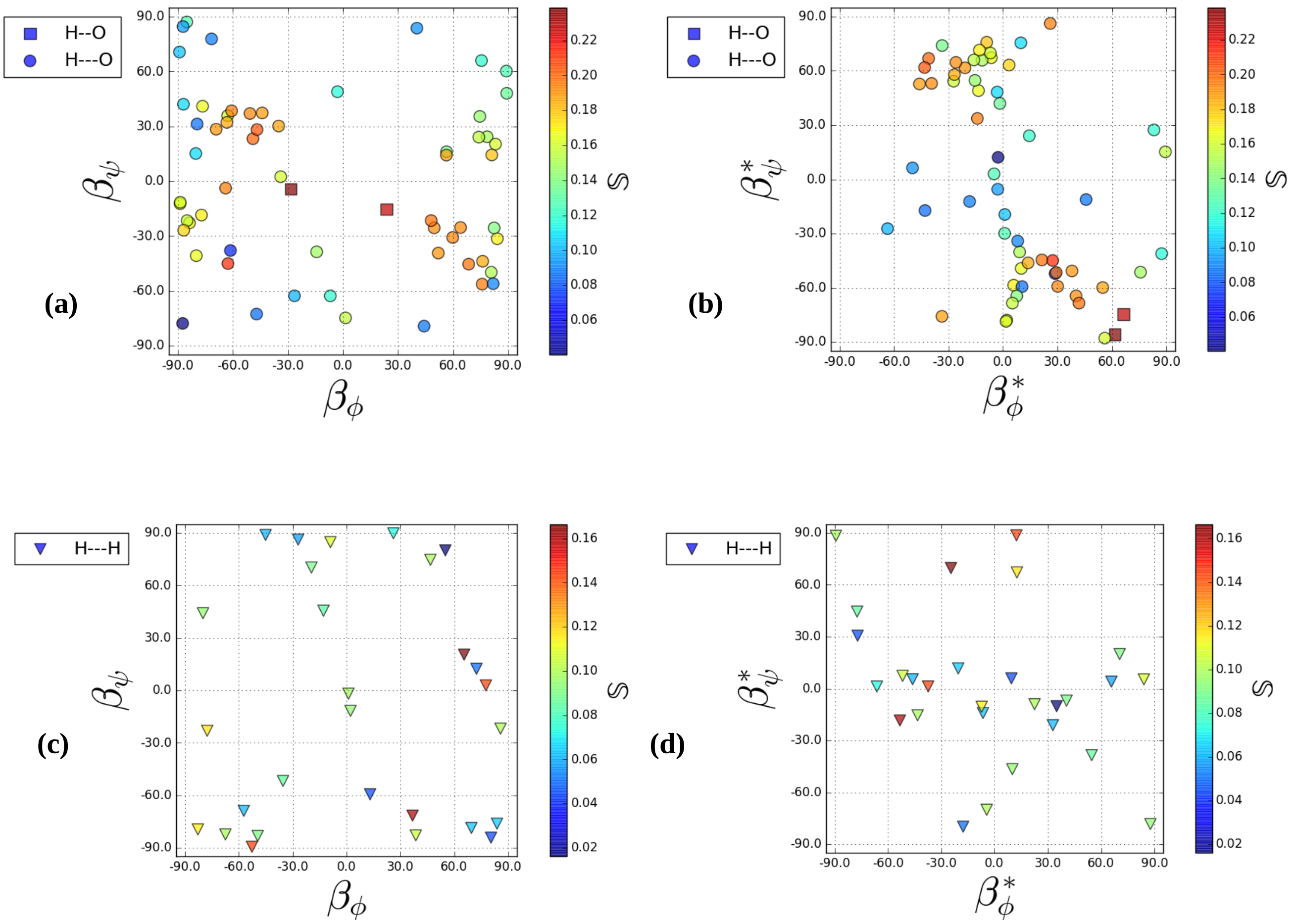

**Figure S3(i).** Plots of the response $\beta_\phi$ versus $\beta_\psi$ with the distribution of stiffness $\mathbb{S}$ of the H--O *BCP*s, H---O *BCP*s and H---H *BCP*s, for the molecular graph of lycosin-01 are shown in sub-figures **(a)** and **(c)** respectively. Plots of the responses $\beta^*_\phi$ versus $\beta^*_\psi$ with the distribution of stiffness $\mathbb{S}$ of the H--O/H---O *BCPs* and H---H *BCPs*, are shown in sub-figures **(b)** and **(d)** respectively. The responses ($\beta_\phi$, $\beta_\psi$) and ($\beta^*_\phi$, $\beta^*_\psi$) indicate the most and least preferred eigenvector responses to structural distortions, see **Scheme 3** for the explanation of the construction of the ($\beta_\phi$, $\beta_\psi$) and ($\beta^*_\phi$, $\beta^*_\psi$).

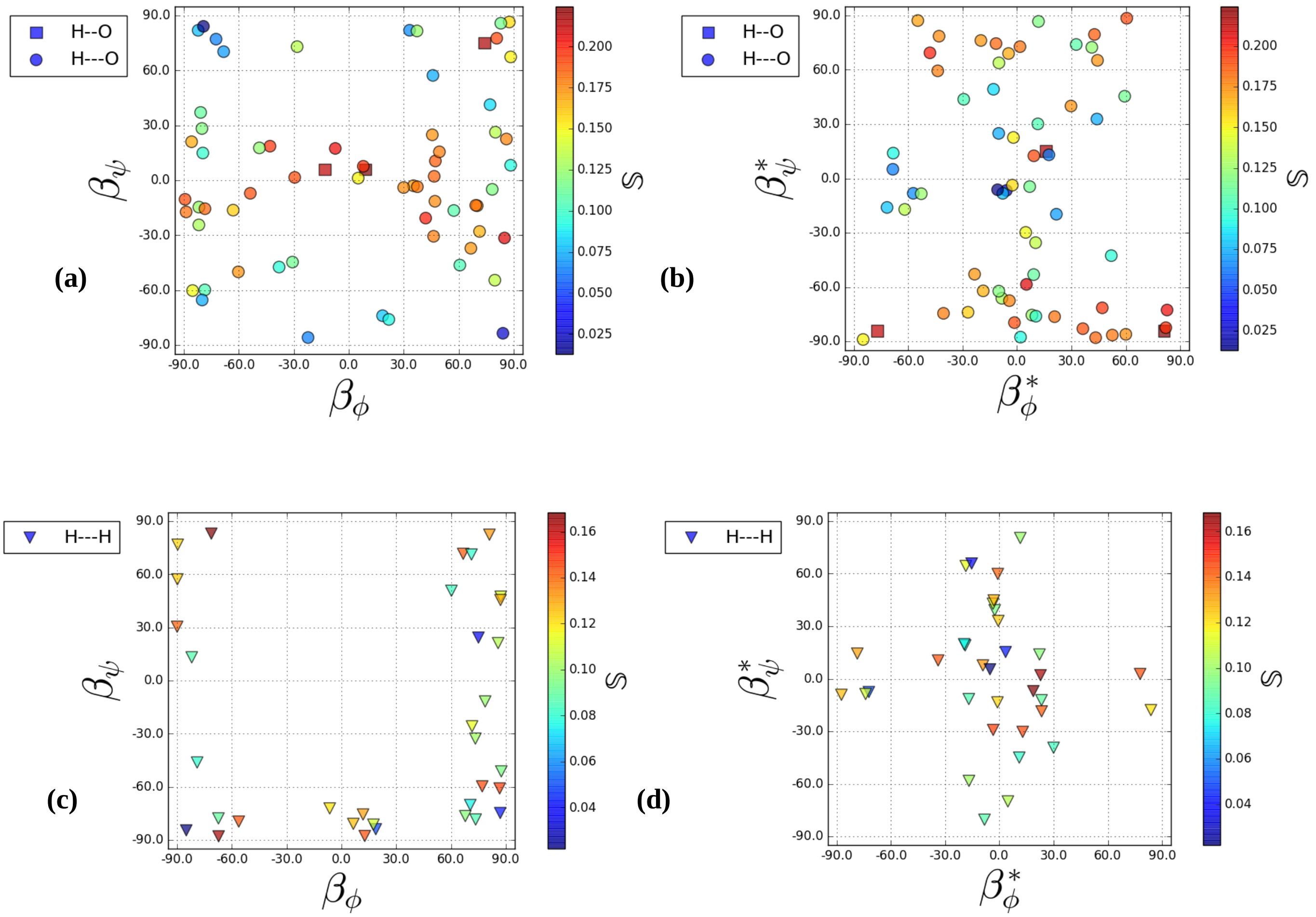

**Figure S3(ii).** Plots of the response $\beta_{\phi}$ versus $\beta_{\psi}$ with the distribution of stiffness $\mathbb{S}$ of the H--O *BCP*s, H---O *BCP*s and H---H *BCP*s for the molecular graph of lycosin-02 are shown in sub-figures **(a)** and **(c)** respectively. Plots of the responses $\beta^{*}_{\phi}$ versus $\beta^{*}_{\psi}$ with the distribution of stiffness $\mathbb{S}$ of the H--O/H---O *BCPs* and H---H *BCPs*, are shown in sub-figures **(b)** and **(d)** respectively.

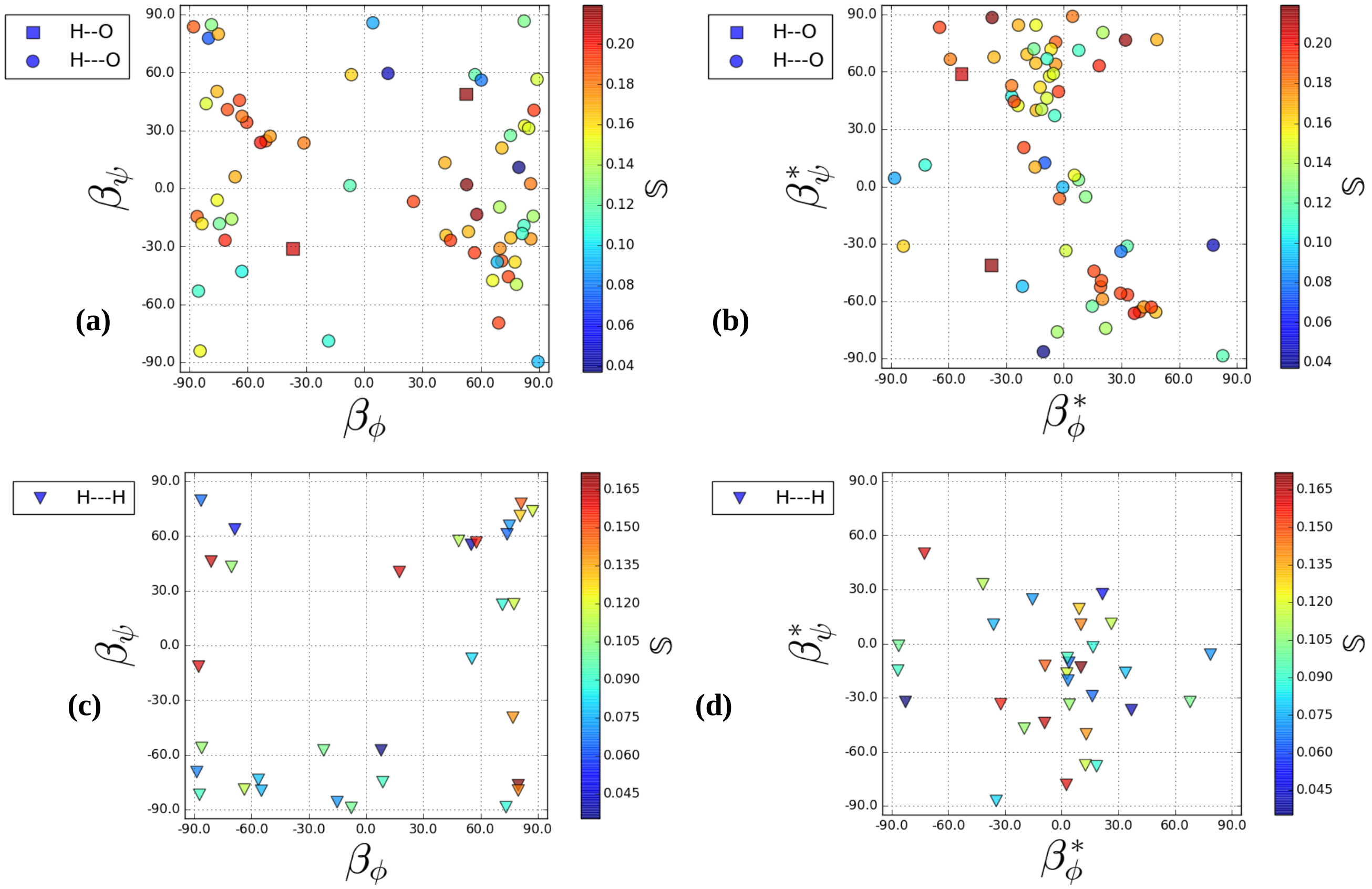

**Figure S3(iii).** Plots of the response $\beta_\phi$ versus $\beta_\psi$ with the distribution of stiffness $\mathbb{S}$ of the H--O *BCP*s, H---O *BCP*s and H---H *BCP*s for the molecular graph of lycosin-03 are shown in sub-figures **(a)** and **(c)** respectively. Plots of the responses $\beta^*_\phi$ versus $\beta^*_\psi$ with the distribution of stiffness $\mathbb{S}$ of the H--O/H---O *BCPs* and H---H *BCPs*, are shown in sub-figures **(b)** and **(d)** respectively.

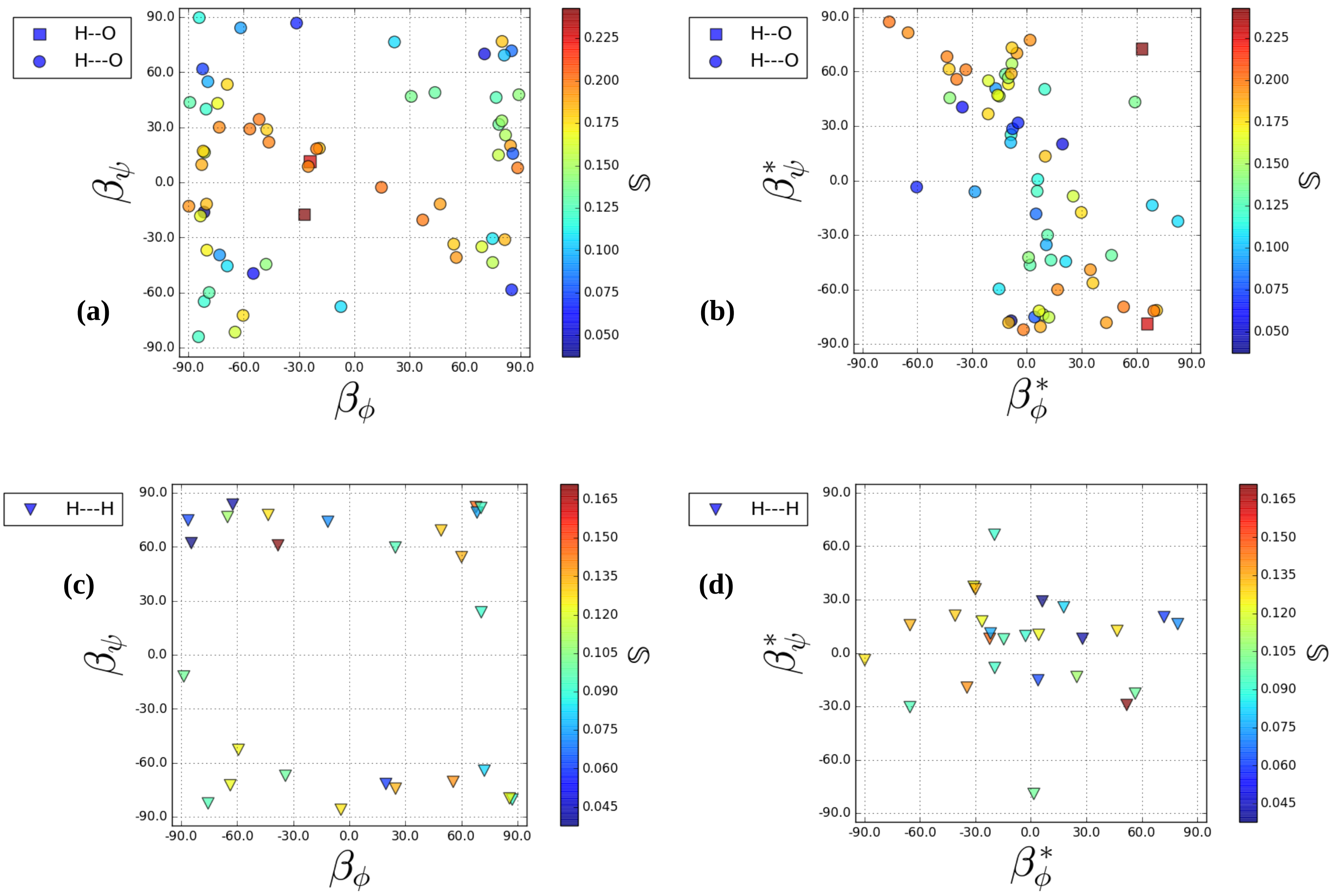

**Figure S3(iv).** Plots of the response $\beta_\phi$ versus $\beta_\psi$ with the distribution of stiffness $\mathbb{S}$ of the H--O *BCP*s, H---O *BCP*s and H---H *BCP*s for the molecular graph of lycosin-04 are shown in sub-figures **(a)** and **(c)** respectively. Plots of the responses $\beta^*_\phi$ versus $\beta^*_\psi$ with the distribution of stiffness $\mathbb{S}$ of the H--O/H---O *BCPs* and H---H *BCPs*, are shown in sub-figures **(b)** and **(d)** respectively.

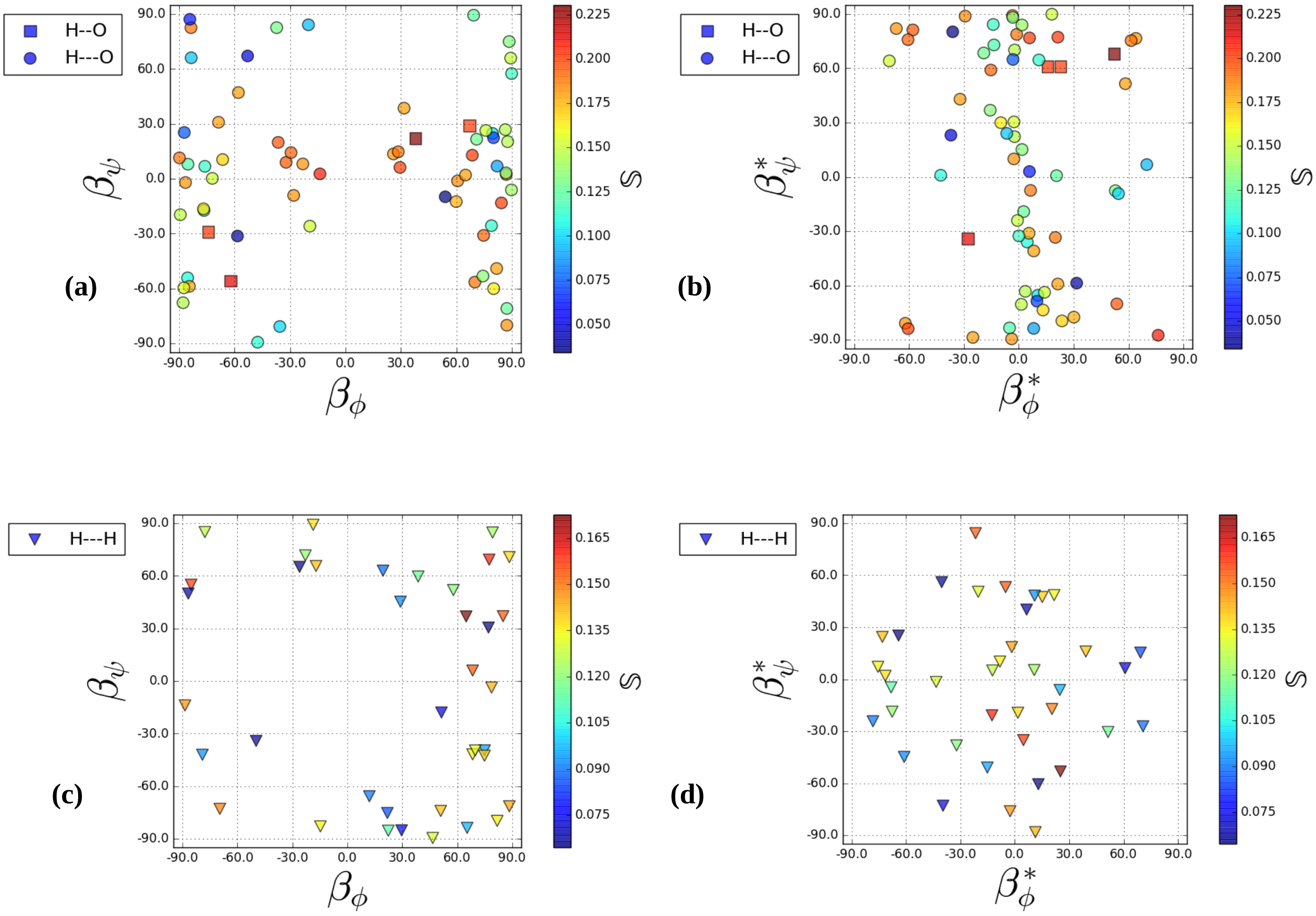

**Figure S3(v).** Plots of the response $\beta_\phi$ versus $\beta_\psi$ with the distribution of stiffness $\mathbb{S}$ of the H--O *BCP*s, H---O *BCP*s and H---H *BCP*s for the molecular graph of lycosin-05 are shown in sub-figures **(a)** and **(c)** respectively. Plots of the responses $\beta^*_\phi$ versus $\beta^*_\psi$ with the distribution of stiffness $\mathbb{S}$ of the H--O/H---O *BCPs* and H---H *BCPs*, are shown in sub-figures **(b)** and **(d)** respectively.

**4. Supplementary Materials S4.** Plots of the responses ($\beta_{\phi}$, $\beta_{\psi}$) and ($\beta^{*}_{\phi}$, $\beta^{*}_{\psi}$) versus the distribution of stiffness $\mathbb{S}$ of the back-bone shared-shell *BCP*s for the five lycosin molecular graphs lycosin-01, lycosin-02, lycosin-03, lycosin-04 and lycosin-05 peptide structures are presented in sub-figures **S4(i)-S4(v)** respectively.

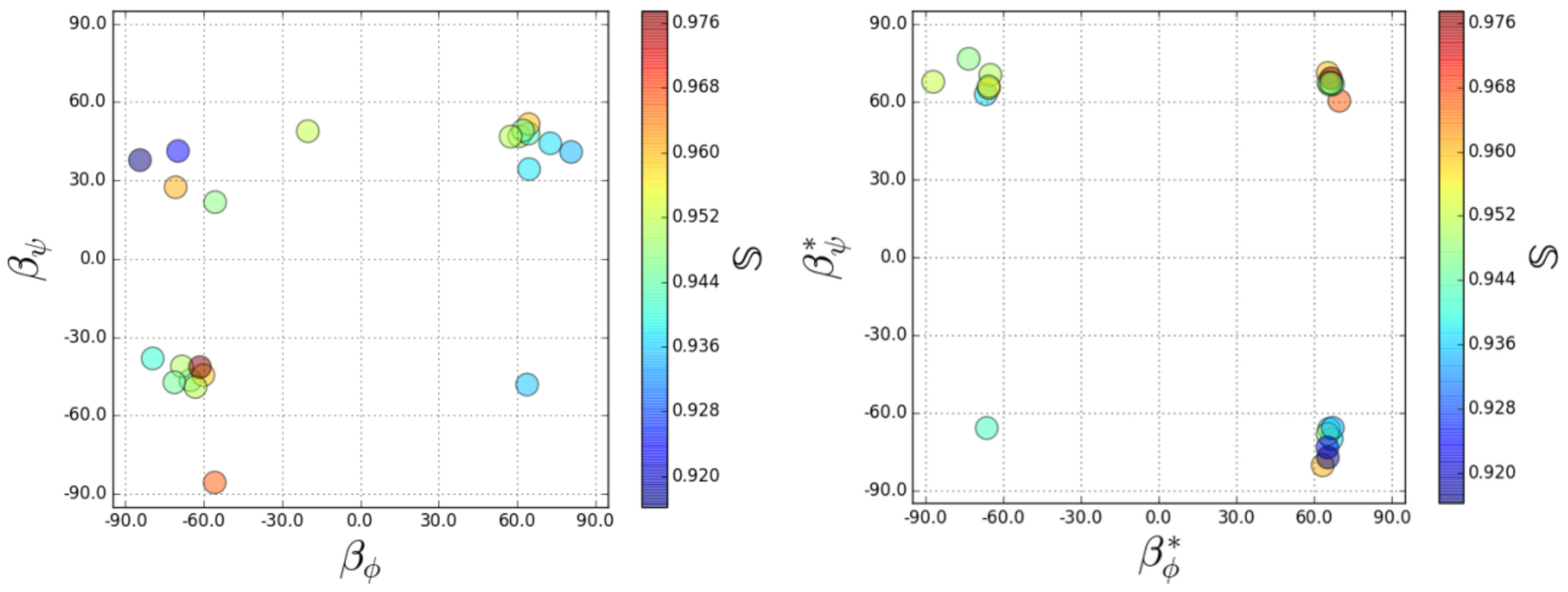

**(a)** **(b)**

**Figure S4(i).** Plots of the response $\beta_{\phi}$ versus $\beta_{\psi}$ with the distribution of stiffness $\mathbb{S}$ and the response $\beta^{*}_{\phi}$ versus $\beta^{*}_{\psi}$ with the distribution of stiffness $\mathbb{S}$ of the backbone shared-shell *BCP*s for the molecular graph of lycosin-01 are shown in sub-figures **(a)** and **(b)** respectively.

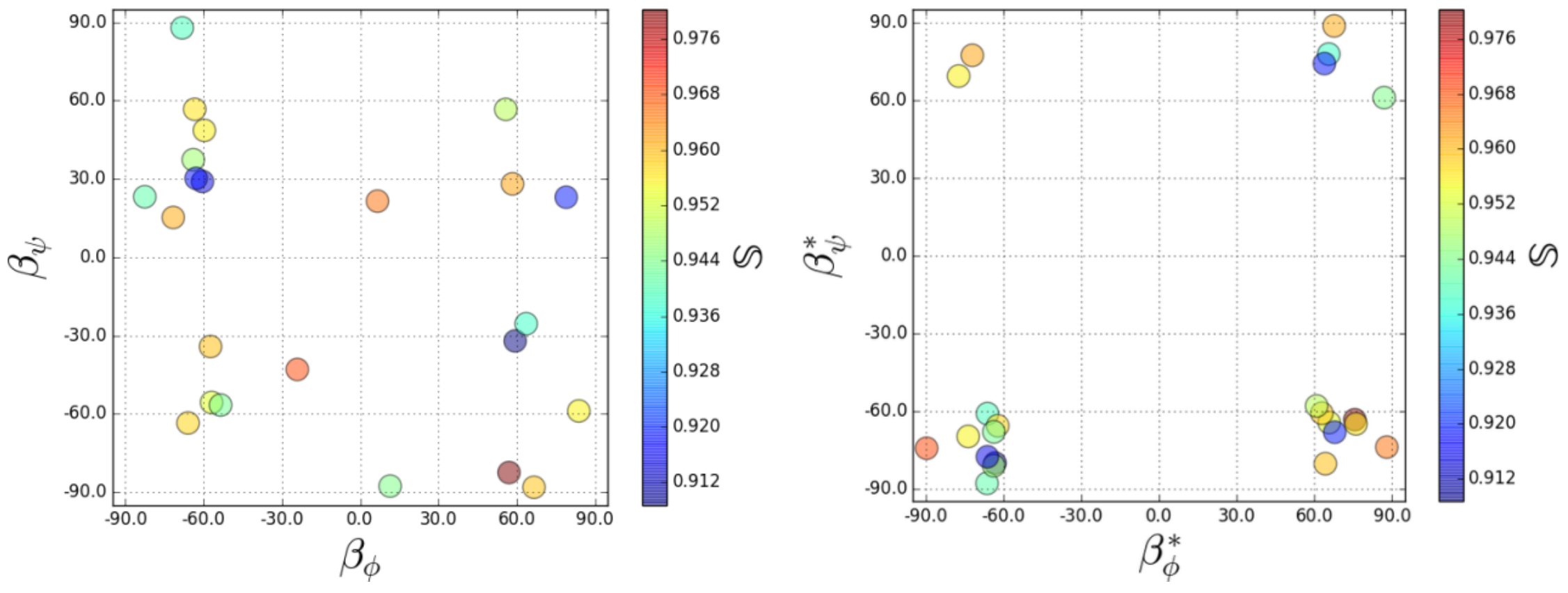

**(a)** **(b)**

**Figure S4(ii).** Plots of the $\beta_{\phi}$ versus $\beta_{\psi}$ with the distribution of $\mathbb{S}$ and the $\beta^{*}_{\phi}$ versus $\beta^{*}_{\psi}$ with the distribution of $\mathbb{S}$ of the backbone shared-shell *BCP*s for the molecular graph of lycosin-02 are shown in sub-figures **(a)** and **(b)** respectively.

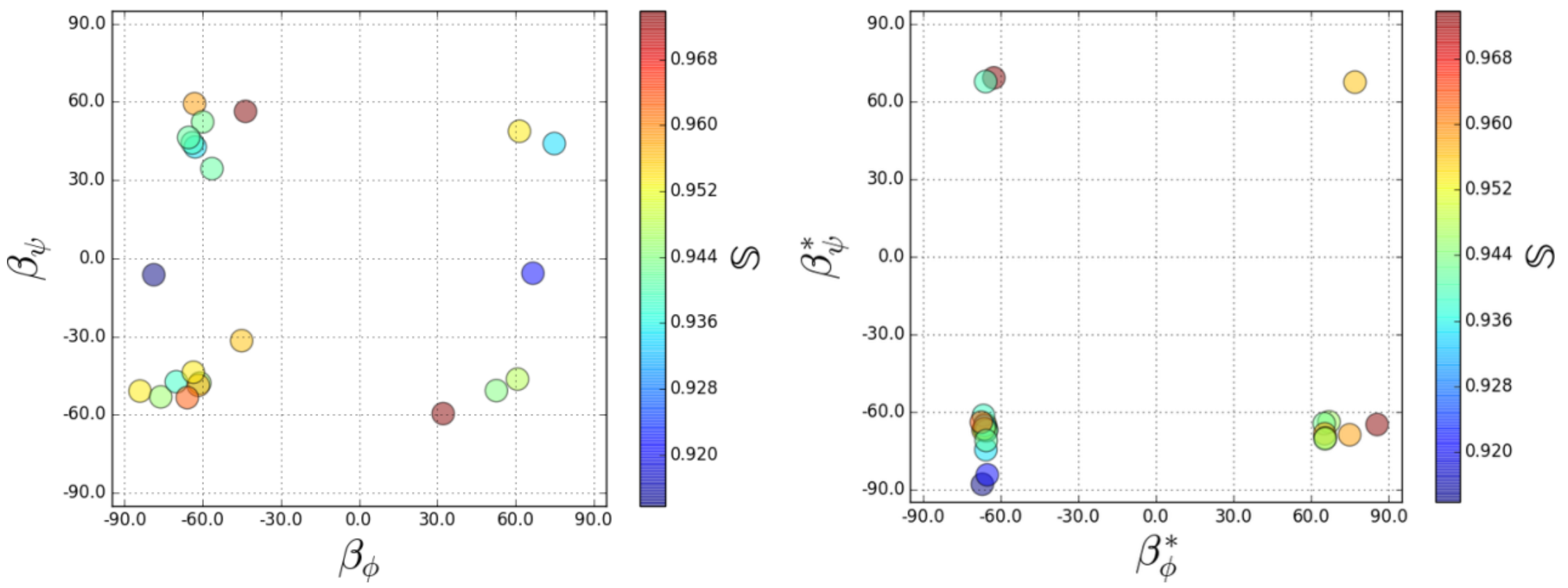

**(a)** **(b)**

**Figure S4(iii).** Plots of the $\beta_\phi$ versus $\beta_\psi$ with the distribution of $\mathbb{S}$ and the $\beta^*_\phi$ versus $\beta^*_\psi$ with the distribution of $\mathbb{S}$ of the backbone shared-shell *BCP*s for the molecular graph of lycosin-03 are shown in sub-figures **(a)** and **(b)** respectively.

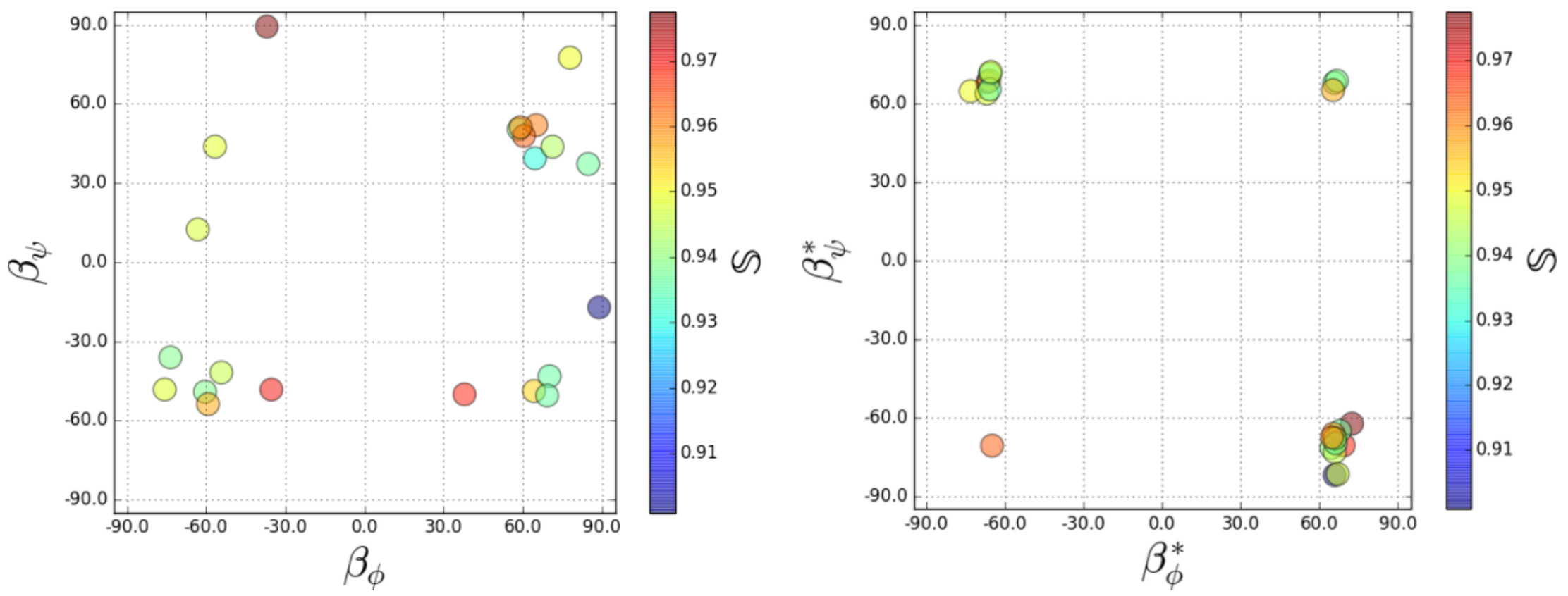

**(a)** **(b)**

**Figure S4(iv).** Plots of the $\beta_\phi$ versus $\beta_\psi$ with the distribution of $\mathbb{S}$ and the $\beta^*_\phi$ versus $\beta^*_\psi$ with the distribution of $\mathbb{S}$ of the shared-shell *BCP*s for the molecular graph of lycosin-04 are shown in sub-figures **(a)** and **(b)** respectively.

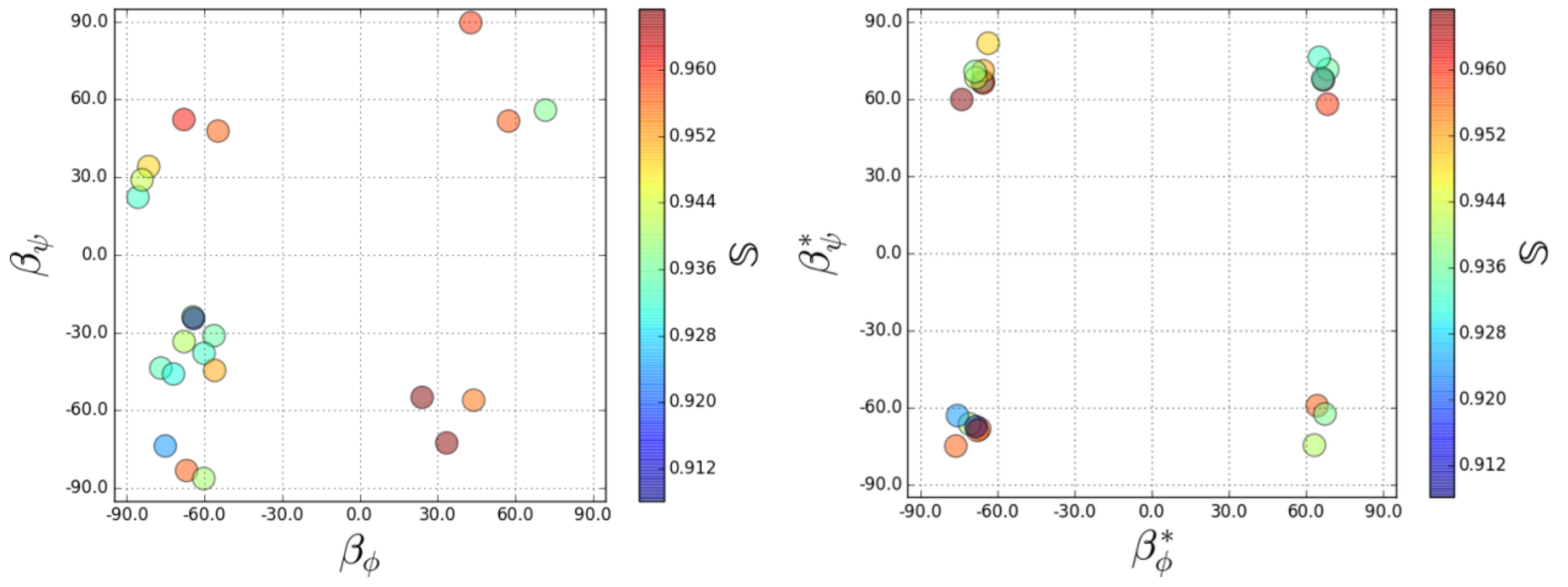

**(a)** **(b)**

**Figure S4(v).** Plots of the $\beta_\phi$ versus $\beta_\psi$ with the distribution of $\mathbb{S}$ and the $\beta^*{}_\phi$ versus $\beta^*{}_\psi$ with the distribution of $\mathbb{S}$ of the backbone shared-shell *BCP*s for the molecular graph of lycosin-05 are shown in sub-figures **(a)** and **(b)** respectively.